\documentclass[twocolumn]{aastex701}

\tabletypesize{\scriptsize}

\received{28-Jul, 2025}
\revised{10-Oct, 2025}
\accepted{16-Oct, 2025}
\submitjournal{ApJS}

\shorttitle{A Spectroscopic Analysis of  Stars with Confirmed Exoplanets from K2 mission}
\shortauthors{Loaiza-Tacuri et al.}

\graphicspath{{./}{figures/}}

\begin{document}

\title{Stellar characterization, Magnesium Abundances and Chromospheric Activity Analysis of Stars with Confirmed Exoplanets from the K2 mission}

\correspondingauthor{Ver\'onica Loaiza Tacuri}

\author[orcid=0000-0003-0506-8269]{V. Loaiza-Tacuri}
\affiliation{Departamento de F\'isica, Universidade Federal de Sergipe, Av. Marcelo Deda Chagas, S/N Cep 49.107-230, S\~ao Crist\'ov\~ao, SE, Brazil}
\affiliation{Observat\'orio Nacional, Rua General Jos\'e Cristino, 77, 20921-400 S\~ao Crist\'ov\~ao, Rio de Janeiro, RJ, Brazil}
\email[show]{vloatac@academico.ufs.br}

\author[orcid=0000-0002-7883-5425]{Diogo Souto}
\affil{Departamento de F\'isica, Universidade Federal de Sergipe, Av. Marcelo Deda Chagas, S/N Cep 49.107-230, S\~ao Crist\'ov\~ao, SE, Brazil}
\email{}

\author[orcid=0000-0001-8741-8642]{F. Quispe-Huaynasi}
\affiliation{Observat\'orio Nacional, Rua General Jos\'e Cristino, 77, 20921-400 S\~ao Crist\'ov\~ao, Rio de Janeiro, RJ, Brazil}
\email{}

\author[orcid=0000-0001-6476-0576]{Katia Cunha}
\affiliation{Steward Observatory, University of Arizona, 933 North Cherry Avenue, Tucson, AZ 85721, USA}
\affiliation{Observat\'orio Nacional, Rua General Jos\'e Cristino, 77, 20921-400 S\~ao Crist\'ov\~ao, Rio de Janeiro, RJ, Brazil}
\email{}

\author[orcid=0000-0001-9205-2307]{S. Daflon}
\affiliation{Observat\'orio Nacional, Rua General Jos\'e Cristino, 77, 20921-400 S\~ao Crist\'ov\~ao, Rio de Janeiro, RJ, Brazil}
\email{}

\author[orcid=0000-0002-1549-626X]{Ellen Costa-Almeida}
\affiliation{Observat\'orio Nacional, Rua General Jos\'e Cristino, 77, 20921-400 S\~ao Crist\'ov\~ao, Rio de Janeiro, RJ, Brazil}
\email{}

\author[orcid=0000-0002-0134-2024]{V. V. Smith}
\affil{NSF’s NOIRLab, 950 N. Cherry Ave. Tucson, AZ 85719 USA}
\email{}

\author[orcid=0000-0002-9089-0136]{Luan Ghezzi}
\affiliation{Universidade Federal do Rio de Janeiro, Observatório do Valongo, Ladeira do Pedro Antônio, 43, Rio de Janeiro, RJ 20080-090, Brazil}
\email{}

\begin{abstract}
We present a homogeneous spectroscopic analysis of confirmed K2 mission exoplanet-hosting stars, comprising 301 targets with high-resolution optical spectra from HIRES and TRES taken from ExoFOP. We derived effective temperatures, surface gravities, and iron and magnesium abundances in LTE by measuring the equivalent widths of Fe I, Fe II, and Mg I lines. Three estimates of stellar masses and radii were obtained via Stefan–Boltzmann and isochrone methods using the codes \texttt{PARAM} and \texttt{isochrones}. These were used to derive exoplanetary radii reaching internal precisions of 2.5\%, 2.6\%, and 6.6\%, respectively, and the radius gap being consistently detected near $1.9 R_{\oplus}$. We measured chromospheric activity from the Ca II H \& K and H$\alpha$ lines. Within the low-activity range ($\log R^{\prime}_{HK}<-4.75$), stellar activity appears to decrease with increasing planetary radius from super-Earths, sub-Neptunes, sub-Saturns, into the Jupiter regime. According to the [Mg/Fe] measurements, most of our K2 planet hosts belong to the Galactic thin disk, but our sample has a population from the thick disk (high-alpha sequence). Most stars show consistent chemo-dynamical behavior. We find that the [Mg/Fe] ratios are indistinguishable between systems containing Large or Small exoplanets, as well as Single- or Multi-exoplanetary systems. Both the [Fe/H] and [Mg/H] distributions reveal that stars hosting large planets are more iron- and magnesium-enhanced than those having only small planets, further confirming the link between stellar abundances and exoplanetary size, but no significant differences are found between the Single- versus Multi-exoplanetary systems. 
\end{abstract}

\section{Introduction} \label{sec:intro}
The discovery and characterization of exoplanets have undergone a significant transformation with the advent of the NASA space-based missions such as Kepler \citep{borucki2010,borucki2016}, K2 \citep{howell2014}, and TESS \citep{ricker2015}. These missions have identified thousands of exoplanet candidates, requiring extensive follow-up observations to confirm their planetary nature and refine their orbital and physical parameters.

Among them, the extended Kepler mission, known as K2 \citep{howell2014}, was the continuation of NASA's original Kepler mission \citep{borucki2010,borucki2016,koch2010}. Following the failure of two of its reaction wheels in 2013, the spacecraft was reoriented to observe different fields along the ecliptic, allowing for the detection of exoplanets and the study of a variety of astronomical objects. 
Notably, K2 targeted a broader and more diverse sample of planet-hosting stars compared to its predecessor, providing new opportunities for demographic and comparative studies. 

Accurate characterization of these stars is essential for determining planetary properties such as radius, which depends directly on the host star's radius. Stellar radii can be estimated using the Stefan-Boltzmann law \citep[e.g.,][]{fulton2018,Martinez2019,Hardegree2020}, isochrone fitting \citep[][]{ghezzi2015ApJ...812...96G,brewer2016a,johnson2017,Brewer2018,Ghezzi2018,mayo2018}, asteroseismology \citep{huber2016ApJS..224....2H}, and interferometry \citep{boyajian2013}.

Accurate stellar radius measurements have been crucial for characterizing exoplanets and have led to significant discoveries, such as the identification of the planetary radius gap \citep{fulton2017,vaneylen2018,Martinez2019}—a noticeable lack of exoplanets with radii between approximately 1.5 and 2 times that of Earth. This gap, also known as the radius valley, exhibits a dependence on the orbital period. That is to say, its location shifts toward smaller radii at longer periods \citep[e.g.,][]{vaneylen2018,Martinez2019}. This feature is commonly attributed to two main mechanisms: photoevaporation driven by mass loss, caused by high-energy stellar radiation stripping planetary atmospheres \citep{owen2012MNRAS.425.2931O,lopez2013,owen2013,owen2017}, and core-powered mass loss, in which the planet’s internal heat drives atmospheric escape \citep{ginzburg2016,ginzburg2018,gupta2019,gupta2020MNRAS.493..792G}. Several works have investigated whether the location and prominence of the radius gap depend on stellar properties, particularly stellar mass. For instance, \citet{fulton2018, berger2020a, cloutier2020, vaneylen2021} reported that the gap's position shifts with stellar mass, supporting a mass-dependent sculpting of planetary atmospheres. Notably, \citet{cloutier2020} demonstrated that the gap remains present even around low-mass stars ($M_\star = 0.08$–$0.93 M_\odot$). In contrast, \citet{petigura2022} analyzed a broader stellar mass range ($M_\star = 0.5$–$1.4 M_\odot$) and found no compelling evidence for a mass dependence of the radius gap, highlighting ongoing uncertainties in how stellar properties shape planet populations.

Recent studies have expanded this picture by examining the behavior of the radius gap across different stellar types. While in F-, G-, and K-type stars the radius gap decreases with orbital period, following a negative slope of $-0.09$ to $–0.11$ \citep{vaneylen2018,Martinez2019}, a different trend emerges among M-dwarf systems. In these low-mass stellar hosts, the radius gap appears to remain nearly constant at $1.8 \pm 0.2$ R$_\oplus$ over a wide range of orbital periods  \citep[$\sim 1-20$ days;][]{wanderley2025arXiv250901930W}. This contrast suggests that the mechanisms governing atmospheric mass loss and planetary evolution may exhibit significant variations depending on the stellar type, possibly attributable to differences in stellar luminosity, spectral energy distribution, and planetary formation conditions \citep{lopez2018MNRAS.479.5303L,gupta2019,luque2022Sci...377.1211L,gaidos2024MNRAS.534.3277G}.

The stellar chromospheric activity is also a crucial parameter in exoplanet studies, as it can influence the determination of stellar parameters and, consequently, planetary properties \citep{yana2019MNRAS.490L..86Y,spina2020ApJ...895...52S}. Stellar activity may introduce spurious signals that can hinder the detection and characterization of planets in both the transit and radial velocity (RV) methods. In the transit method, spots and faculae on the stellar surface cause quasi-periodic brightness modulations linked to rotation, complicating the detection of small-radius planets \citep{pont2013MNRAS.432.2917P,rackham2018ApJ...853..122R}. Transits across active regions may alter the apparent depth depending on the contrast with the surrounding photosphere \citep{oshagh2013AA...556A..19O}. Flares and other variability increase photometric noise, especially in active M dwarfs \citep{davenport2016ApJ...829...23D}.
In RV measurements, surface inhomogeneities distort spectral lines as the star rotates, inducing RV shifts unrelated to orbital motion \citep{saar1997ApJ...485..319S,queloz2001AA...379..279Q,boisse2011AA...528A...4B}. These activity-driven signals can mimic or mask those of Earth-mass planets, particularly around young or active stars \citep{dumusque2012Natur.491..207D,lovis2011}.
Stellar activity may also affect planetary habitability, especially around cool stars where intense flares, UV/X-ray radiation, and stellar winds can erode atmospheres and alter surface conditions \citep{segura2018haex.bookE..73S}.

To better understand and mitigate these effects, especially in the context of planet detection and characterization, several stellar activity indicators are used in the literature to trace and correct for the effects of stellar activity on radial velocity measurements \citep[e.g.;][]{boisse2011AA...528A...4B,dumusque2012Natur.491..207D}. Among the most widely adopted optical diagnostics are the Ca II H \& K and H$\alpha$ \citep[e.g.,][]{cinecungui2007AA...469..309C,boisse2009AA...495..959B,isaacson2010ApJ...725..875I,gomes2011AA...534A..30G,gomes2014AA...566A..66G,gomes2021AA...646A..77G, boro_saikia2018AA...616A.108B,ibanez2023AA...672A..37I}. These indicators are often used in conjunction with each other, under the assumption that they measure similar aspects of chromospheric activity.
The correlation between the Ca II H \& K and H$\alpha$ indices is not yet fully understood. While stars with higher chromospheric activity levels often show a strong positive correlation between these indicators, the situation is less clear for stars with lower or intermediate activity. In these cases, studies have reported a broad range of behaviors, including positive, weak, and even negative correlations \citep{cinecungui2007AA...469..309C,santos2010AA...511A..54S,gomes2011AA...534A..30G,gomes2014AA...566A..66G,meunier2022AA...658A..57M}.
This diversity suggests that the two indicators may trace different aspects or layers of chromospheric activity. In particular, a given level of H$\alpha$ absorption can correspond to a wide range of Ca II H \& K emission levels, especially in less active stars \citep{walkowicz2009AJ....137.3297W}. Recent studies based on high-resolution spectroscopy have explored the relationship between chromospheric activity and the fundamental properties of exoplanet-hosting stars. For example, \citet{isaacson2024ApJ...961...85I} analyzed a sample of confirmed Kepler planet hosts using the Ca II H \& K lines as tracers of stellar activity. One of their main findings was the lack of a significant correlation between chromospheric activity and metallicity, at least within the range [Fe/H] $\sim -0.2$ to +0.3 dex. 
A similar result was also reported by \citet{loaiza2024ApJ...970...53L} for a sample of stars hosting exoplanets from the K2 mission, suggesting that this apparent independence between activity and metallicity may hold across different planetary populations observed with optical spectroscopy.

In this study, we analyze high-resolution spectra from ExoFOP \citep[the Exoplanet Follow-up Observing Program;][]{exofop2}, which serves as a centralized repository of stellar and planetary data for confirmed systems. For K2 targets alone, ExoFOP provides publicly available spectroscopic and photometric data for over 500 exoplanet-hosting stars, supporting homogeneous and large-scale analyses. We focus on a sample of 301 confirmed exoplanet-hosting stars discovered by the K2 mission. This represents the largest homogeneous analysis of confirmed K2 exoplanet host stars based on high-resolution optical spectroscopy. We derive stellar parameters through a quantitative spectroscopic analysis using carefully selected Fe I and Fe II lines. By combining these spectroscopic parameters with astrometric and photometric data, we estimate stellar masses and radii using both the Stefan–Boltzmann law and the isochrone method, applying publicly available tools. We also explore how the choice of method affects the inferred planetary radius gap. In addition, we determined chromospheric activity levels using the Ca II H \& K and H$\alpha$ lines to investigate stellar behavior in different activity regimes. Given the diversity of correlations reported between these indicators, ranging from strong positive to absent or even negative, we examine how this variability manifests within our sample. This work builds on a subset of stars previously analyzed in \citet{loaiza2024ApJ...970...53L}, which are now incorporated into a larger sample, enabling a more comprehensive investigation. Our goal is to provide a homogeneous determination of atmospheric parameters for all stars in the sample by extending the spectroscopic analysis to include both iron and magnesium (an alpha-element), key tracers of stellar and Galactic chemical evolution. 
In particular, the study of [Mg/Fe] ratios as a function of [Fe/H] allows us to segregate the exoplanet hosting stellar sample into chemically belonging to either the thin or thick disk, while their dynamics will also be explored in this study.

This paper is organized as follows. Section \ref{sec:data} describes the observation data. Section \ref{sec:st_par} discusses the methodology used to derive key stellar parameters and uncertainties, including effective temperatures ($T_{\rm eff}$), surface gravities (log $g$), iron abundances (which can be used as proxies for metallicities), magnesium abundances, stellar masses, radii, and planetary radii. In Section \ref{sec:S_ind} we explain the approach used to assess stellar activity using the Ca II H \& K lines and the H$\alpha$ lines. Sections \ref{sec:discus} and \ref{sec:conclusion} present discussions and conclusions, respectively.

\section{Data} \label{sec:data}
The sample consists of stars with confirmed exoplanets observed by the Kepler Extended Mission (K2) during the C0--C8 and C10--C18 campaigns, whose high-resolution optical spectra were obtained from the ExoFOP database. These spectra were acquired using two instruments: the High-Resolution Echelle Spectrometer \citep[HIRES;][]{vogt1994} on the Keck I 10m telescope and the Tillinghast Reflector Echelle Spectrograph \citep[TRES;][]{furesz2008} on the 1.5m Tillinghast telescope at the Fred Lawrence Whipple Observatory on Mt. Hopkins, Arizona.

The HIRES spectra, acquired by the California Planet Search (CPS) program after August 2004 with upgraded CCDs, cover the 3640–7990 \text{\r{A}} range and include key activity indicators such as the Ca II H \& K lines and H$\alpha$. A total of 2320 reduced HIRES spectra corresponding to 198 stars were obtained from ExoFOP. A subsequent inspection of these spectra revealed that 165 of them possessed signal-to-noise ratios sufficiently high to facilitate analysis to derive stellar parameters and metallicities (Section \ref{sec:st_par}). 
The TRES spectra span the 3800–9100 \text{\r{A}} range and contribute 160 spectra of 103 stars.

Finally, we have a total of 301 stars whose observational data, such as identifiers, positions, B and V magnitudes (taken from the EPIC catalog and the NASA Exoplanet Archive), K2 observing campaigns, spectrograph information (HIRES or TRES), and the number of spectra of the sample stars are given in Table \ref{tab:sample}.

Among the spectra of confirmed planet-hosting stars obtained from ExoFOP, we also have stars classified as M-type according to the SIMBAD Astronomical Database. For these stars, only stellar activity will be determined, as the method used to estimate stellar parameters is not suitable for these types of stars. 

\startlongtable
\begin{deluxetable*}{llccccccc}
\tablecaption{Sample K2 Stars \label{tab:sample}}
\tablecolumns{7}
\tablenum{1}
\tablewidth{0pt}
\tablehead{
\colhead{ID} & \colhead{Host Name} & \colhead{R.A.} & \colhead{Decl.} & \colhead{$B$} & \colhead{$V$} & \colhead{Camp} & \colhead{Spectrograph} & \colhead{N$\rm _{spec}$} \\
\colhead{} & \colhead{} & \colhead{(deg)} & \colhead{(deg)} & \colhead{(mag)} & \colhead{(mag)} & \colhead{} & \colhead{} & \colhead{}
}
\startdata
EPIC201295312 & K2-44 & 174.0116610 & -2.5209338 & 12.782 & 12.189 & 1 & HIRES & 1 \\
EPIC201367065 & K2-3 & 172.3353708 & -1.4551364 & 13.524 & 12.167 & 1 & HIRES & 146 \\
EPIC201384232 & K2-6 & 178.1921152 & -1.1984418 & 13.302 & 12.660 & 1 & HIRES & 1 \\
EPIC201403446 & K2-46 & 174.2662454 & -0.9072100 & 12.485 & 12.029 & 1 & HIRES & 1 \\
EPIC201445392 & K2-8 & 169.7935178 & -0.2844389 & 15.726 & 14.609 & 1 & HIRES & 1 \\
\nodata & \nodata & \nodata & \nodata & \nodata & \nodata & \nodata & \nodata & \nodata \\ 
\enddata
\tablecomments{This table is published in its entirety in the machine-readable format. A portion is shown here for guidance regarding its form and content.}
\end{deluxetable*}

\section{Stellar Parameters and Abundances} \label{sec:st_par}

\subsection{Effective temperatures, Surface Gravities, Iron and Magnesium Abundances}

The stellar parameters of $T_{\rm eff}$, $\log g$, iron abundance (A(Fe)\footnote{A(X)=log(N(X)/N(H))+12.0}), and microturbulent velocity ($\xi$) were derived by enforcing the excitation/ionization equilibrium from a selected set of Fe I and Fe II spectral lines with equivalent width measurements. 
The analysis assumes local thermodynamic equilibrium (LTE) and uses 1-D plane-parallel model atmospheres from the Kurucz ATLAS9 ODFNEW grid \citep{castelli2004a}.

The line list used in this study was taken from \cite{Ghezzi2018,Ghezzi2021} and contains 133 Fe I and 18 Fe II lines. Table \ref{tab:linelist} lists the wavelengths of the Fe I (species=26.0) and Fe II (species=26.1) lines along with their excitation potentials and $\log$ $gf$ values (solar). 
We used the ARES code v2 \citep{sousa2015A&A...577A..67S} to set the continuum and measure the EWs of all Fe lines in this study. Three line-free spectral regions, $\lambda$ 5764 - 5766 \AA, $\lambda$ 6047 - 6052 \AA, and $\lambda$ 6068 - 6076 \AA, were used to set the continuum level.

The spectroscopic analysis performed in this work follows a methodology similar to that applied in our previous studies of K2 targets \citep{loaiza2023ApJ...946...61L,loaiza2024ApJ...970...53L}. Specifically, we use an automated code, \texttt{qoyllur-quipu4} (q2; \citealt{Ramirez2014}), designed to derive stellar parameters and metallicities.
\textsc{$q^2$} uses our predefined list of iron lines with measured equivalent widths and works in conjunction with the 2019 version of the abundance analysis software \texttt{MOOG} (\citealt{Sneden1973}) to determine simultaneously iron abundances, effective temperatures, surface gravities, and metallicities.

\begin{figure}
    \centering
    \includegraphics[width=\linewidth]{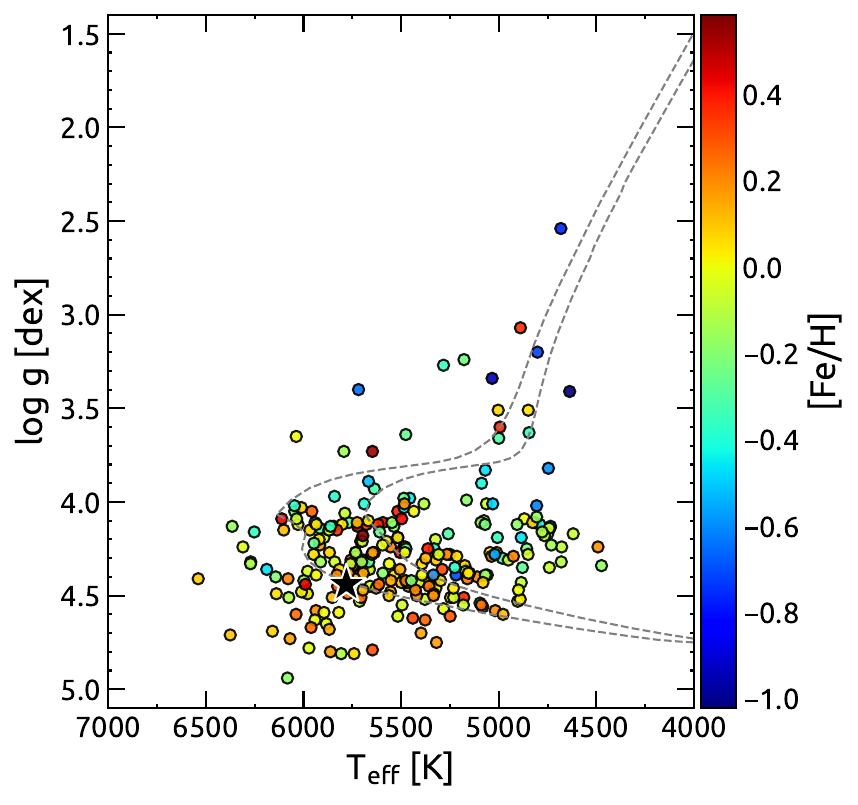}
    \caption{Kiel diagram of the sample stars. Gray dashed lines show solar-metallicity evolutionary tracks for 4.6 and 10 Gyr from the Yonsei–Yale models. The solar proxy result is indicated by a black star.}
    \label{fig:kiel_diagram}
\end{figure}

The procedure begins with an initial atmospheric model, interpolated using assumed values of $T_{\rm eff}$, $\log g$, and metallicity as an initial guess. These parameters are refined iteratively by adjusting $T_{\rm eff}$ / $\log g$ / A(Fe) to eliminate trends in A(Fe I) and A(Fe II) as functions of $T _{\rm eff}$, $\log g$, and $\xi$. This iterative approach continues until a self-consistent solution is reached, yielding the final spectroscopic parameters for each star. This is a classical method widely used in the literature for obtaining atmospheric parameters using optical spectra \citep[e.g.,][]{Martinez2019, Ghezzi2021}.
The effective temperatures, surface gravities, metallicities, and microturbulent velocities derived for all analyzed stars are presented in Table \ref{tab:parameters}. We show the stellar parameters of $\rm T_{eff}$ and $\log g$ for the K2 stars in this study in the Kiel diagram presented in Figure \ref{fig:kiel_diagram}. Filled circles represent the target stars, color-coded according to their metallicities, as indicated by the color bar. We also show two Yonsei–Yale isochrones \citep{yi2001,yi2003ApJS..144..259Y,demarque2004ApJS..155..667D,han2009gcgg.book...33H} corresponding to solar metallicity for 4.6 and 10 Gyr as dashed gray lines.
As a test of our methodology, we analyzed solar proxy spectra obtained with the HIRES spectrograph by targeting the reflected sunlight from the asteroid Iris. We also included the solar parameters derived here in Table \ref{tab:parameters}. 
The results are generally consistent with the known solar parameters, suggesting that our method yields reliable estimates for solar-type stars and is not significantly affected by systematic biases. The mean iron abundance obtained from our solar proxies (A(Fe) = 7.52) is slightly higher than the value reported by \citet[][A(Fe) = 7.46]{asplund2021AA...653A.141A}, but it agrees well with the value reported by \citet[][A(Fe) = 7.50]{magg2022AA...661A.140M}.

We determined magnesium abundances for all stars in our sample, using the derived stellar parameters and the equivalent width method.  
For this analysis, we selected four Mg~I lines (listed in Table \ref{tab:linelist}). The final magnesium abundance for each star was obtained as the mean of the individual abundances derived from each line. The results are presented in Table \ref{tab:parameters}.

\vspace{-\baselineskip}
\begin{deluxetable}{lcccccccccccc}
\tablecaption{Line List\label{tab:linelist}}
\tablecolumns{12}
\tablenum{2}
\tablewidth{0pt}
\tablehead{
\colhead{$\lambda$} & \colhead{Species} & \colhead{$\chi$} & \colhead{$\log$ $gf$} & \\ 
\colhead{(\AA)} & \colhead{} & \colhead{(eV)} & \colhead{}  & \\ 
}
\startdata
5023.185    &   26.0    &   4.283   &   -1.524   \\
5025.303    &   26.0    &   4.284   &   -1.919   \\
5054.642    &   26.0    &   3.640   &   -2.087   \\
5058.496    &   26.0    &   3.642   &   -2.809   \\
5197.935    &   26.0    &   4.301   &   -1.608   \\
\nodata	&	\nodata	&	\nodata	&	\nodata  	 \\
\enddata
\tablecomments{The complete table containing the iron and magnesium line list is available in a machine-readable format.}
\end{deluxetable}

\begin{deluxetable*}{lllllcccccc}
\tabletypesize{\tiny}  
\setlength{\tabcolsep}{4pt}  
\tablecaption{Stellar Parameters \label{tab:parameters}}
\tablecolumns{11}
\tablenum{3}
\tablewidth{0pt}
\tablehead{
\colhead{ID} & \colhead{$\langle\rm S_{HK}\rangle$} & \colhead{$\langle\rm \log R^\prime_{HK}\rangle$} & \colhead{$\langle\rm H\alpha\rangle$} & \colhead{$\langle\rm \log I_{H\alpha}\rangle$} &  \colhead{$T_{\rm eff}$} & \colhead{$\log g$} & \colhead{A(Fe)} & \colhead{$\xi$} & \colhead{$\rm A(Mg)$} \\
\colhead{} & \colhead{} & \colhead{} & \colhead{} & \colhead{} & \colhead{(K)} & \colhead{(dex)} & \colhead{(dex)} & \colhead{($\rm km~s^{-1})$} & \colhead{(dex)} 
}
\startdata
Sun &   &   &  &   &   5780 $\pm$ 4  & 4.440 $\pm$ 0.011 & 7.522 $\pm$ 0.004 & 0.960 $\pm$ 0.020 & 7.588 $\pm$ 0.029 \\
K2-44 & 0.146$\pm$ ... & -5.038$\pm$ ...  & 0.030$\pm$ ...  & -1.430$\pm$ ... &  $5917 \pm 24$ & $4.230 \pm 0.056$ & $7.713 \pm 0.015$ & $1.350 \pm 0.050$ & $7.730 \pm 0.173$ \\
K2-3 & 0.575$\pm$ 0.054  & -4.680$\pm$0.399    & 0.056$\pm$0.001    & -1.242$\pm$0.026   &  \nodata       & \nodata         & \nodata             & \nodata           & \nodata \\
K2-6 & 0.173$\pm$ ... & -4.918$\pm$ ...  & 0.031$\pm$ ...  & -1.420$\pm$ ... &  $5713 \pm 29$ & $4.530 \pm 0.049$ & $7.441 \pm 0.018$ & $0.870 \pm 0.090$ & $7.300 \pm 0.134$ \\
K2-46 & 0.138$\pm$ ... & -4.923$\pm$ ...  & 0.029$\pm$ ...  & -1.433$\pm$ ... &  $6252 \pm 83$ & $4.160 \pm 0.127$ & $7.161 \pm 0.047$ & $1.180 \pm 0.320$ & $7.291 \pm 0.082$ \\
K2-8 & 0.221$\pm$ ... & -5.086$\pm$ ...  & 0.041$\pm$ ...  & -1.382$\pm$ ... &  $5011 \pm 30$ & $4.270 \pm 0.069$ & $7.498 \pm 0.020$ & $0.890 \pm 0.090$ & $7.552 \pm 0.038$ \\
\nodata & \nodata & \nodata & \nodata & \nodata & \nodata & \nodata & \nodata & \nodata & \nodata  \\
\enddata
\tablecomments{A portion is shown here for guidance regarding its form and content.}
\end{deluxetable*}


\subsection{Stellar Masses \& Radii} \label{sec:radii}
Stellar masses and radii are critical in defining planetary masses and radii. 
We determined these stellar properties from the fundamental equations and the isochrone method. 
Firstly, we derived the stellar radii using a similar methodology as \cite{Martinez2019}, that is, using the Stefan-Boltzmann law (hereafter, we will refer to it as SB), which depends on the Stefan-Boltzmann constant ($\sigma _B$), the stellar effective temperature ($T_{\rm eff}$), and luminosity ($L_\star$)

\begin{equation}
    R_{\star} = \left( \frac{L_{\star}}{4 \pi \sigma _{SB} T_{\text{eff}}^4} \right) ^{1/2},
\end{equation}

where $L_{\star} = L_0 10^{-0.4}M_{bol}$, here $L_0$ represents the zero point of the bolometric magnitude scale \citep{mamajek2015}, while $M_{bol}$ denotes the bolometric magnitude, which is defined as: 

\begin{equation}
    M_{bol} = m_k - A_k - \mu + BC_k,
\end{equation}
where $m_k$ is the photometric apparent magnitude, $A_k$ is the extinction, $BC_k$ is the bolometric correction within the same band, and $\mu$ is the distance modulus. 

To determine $\mu$ for each target star, we adopted the photo-geometric distances from \cite{bailerjones2021}, which uses the parallaxes, the $G$ magnitude, and the $BP-RP$ color from Gaia EDR3 \citep{gaia2021A&A...649A...1G}.
We used 2MASS $K_s-$ band photometry in conjunction with reddening $E(B - V)$, which was obtained from IRSA/Galactic Dust Reddening and Extinction\footnote{https://irsa.ipac.caltech.edu/applications/DUST/}. 
The extinction in the $K$ band ($A_k$) was determined by applying the conversion relations of \citet{Bilir2008}. Additionally, we employed the \texttt{isoclassify} package \citep{Huber2017} in its ``direct mode'', incorporating $T_{\rm eff}$, log $g$, [Fe/H], and $A_V$ as input parameters to interpolate bolometric corrections from the MIST model grids \citep{Choi2016}. This approach allowed us to compute absolute magnitudes and stellar luminosities. 

The other method used in this work to determine the stellar masses and radii was the isochrones method, which involves estimating these fundamental parameters by comparing the positions of stars in a $T_{\rm eff}$ -- $M_V$ diagram with theoretical isochrones. 
To perform this analysis, we used the \texttt{PARAM} code v1.3 \citep{dasilva2006AA...458..609D}, which derives stellar mass, radius, luminosity, age, and other parameters based on a grid of PARSEC isochrones \citep{bressan2012} using a Bayesian inference methodology. 
We also determined stellar masses and radii using the \texttt{isochrones} code \citep{morton2015}. 
This code incorporates several features, including the use of the MESA Isochrones \& Stellar Tracks models \citep[MIST;][]{Choi2016,dotter2016ApJS..222....8D} to infer stellar properties, such as mass, radius, and age, from a set of arbitrary observables \citep{morton2015}. 
The input parameters required by both the \texttt{PARAM} and \texttt{isochrones} codes are $T_{\rm eff}$ and [Fe/H], which were determined spectroscopically in this work, as well as the absolute magnitude ($M_V$), derived using the parallax measurements obtained from the Gaia DR3 data \citep{gaia2021A&A...649A...1G}.
Table \ref{tab:R_M} shows the masses and radii derived from the sample stars.

\begin{deluxetable*}{lccccc}
\tabletypesize{\scriptsize}  
\setlength{\tabcolsep}{2pt}  
\tablecaption{Stellar Masses and Radii\label{tab:R_M}}
\tablecolumns{6}
\tablenum{4}
\tablewidth{0pt}
\tablehead{
\colhead{} & \colhead{SB} & \colhead{\texttt{PARAM}} & \colhead{} & \colhead{\texttt{isochrones}} & \colhead{} \\
\colhead{Star} & \colhead{R$_\star$ (R$_\odot$)} & \colhead{R$_\star$ (R$_\odot$)} & \colhead{M$_\star$ (M$_\odot$)} & \colhead{R$_\star$ (R$_\odot$)} & \colhead{M$_\star$ (M$_\odot$)}
}
\startdata
K2-44 & $1.517\pm0.026$ & $1.487\pm0.033$ & $1.207\pm0.030$ & $1.678^{+0.138}_{-0.162}$ & $1.254^{+0.045}_{-0.046}$ \\
K2-6 & $0.889\pm0.010$ & $0.887\pm0.020$ & $0.967\pm0.019$ & $1.018^{+0.104}_{-0.167}$ & $0.936^{+0.015}_{-0.015}$ \\
K2-46 & $1.365\pm0.051$ & $1.361\pm0.043$ & $1.038\pm0.030$ & $1.514^{+0.121}_{-0.229}$ & $1.113^{+0.048}_{-0.058}$ \\
K2-8 & $0.712\pm0.007$ & $0.714\pm0.015$ & $0.806\pm0.010$ & $0.771^{+0.027}_{-0.026}$ & $0.798^{+0.013}_{-0.013}$ \\
K2-19 & $0.829\pm0.006$ & $0.832\pm0.002$ & $0.942\pm0.012$ & $0.892^{+0.047}_{-0.068}$ & $0.904^{+0.014}_{-0.015}$ \\
\nodata	&	\nodata	&	\nodata	&	\nodata	&	\nodata &	\nodata	 \\
\enddata
\tablecomments{The complete table is available in machine-readable format.}
\end{deluxetable*}

\subsection{Planetary Radii} \label{sec:planet}
We derived the planetary radii from stellar radii and transit depth ($\Delta F$), which represents the fraction of stellar flux lost during the planet's transit minimum, given by \cite{seager2003}:

\begin{equation}
    R_{pl} = 109.1979 \times \sqrt{\Delta F \times 10^{-6}} \times R_{star}, 
\end{equation}
where the planet's radius is given in terms of Earth's radius.

The $\Delta F$ values were obtained from \cite{kruse2019}, \cite{guerrero2021ApJS..254...39G}, \cite{livingston2018}, \cite{christiansen2022AJ....163..244C}, \cite{yu2018AJ....156..127Y}, \cite{mayo2018}, \cite{vanderburg2016}, \cite{adams2021PSJ.....2..152A}, \cite{kokori2023ApJS..265....4K}, \cite{pope2016MNRAS.461.3399P}, \cite{hidalgo2020AA...636A..89H}, \cite{thygesen2023AJ....165..155T}, \cite{david2018AJ....155..222D}, \cite{dattilo2019AJ....157..169D}, \cite{hjorth2019MNRAS.484.3522H}, \cite{barros2016AA...594A.100B}, \cite{mann2017AJ....153...64M},  \cite{diez2019MNRAS.489.5928D}, \cite{jordan2019AJ....157..100J}, \cite{luque2019AA...623A.114L}, \cite{castro2020MNRAS.499.5416C}, \cite{polanski2021AJ....162..238P}, and \cite{sha2021AJ....161...82S}, and these values, along with star identifications, planet names, planetary radii ($\rm R_{pl}$) and their corresponding errors ($\rm \delta R_{pl}$) for all stellar radii derived here, are listed in Table \ref{tab:planet_rad}.
We note that our analysis only considered the $\Delta F$ values of confirmed exoplanets. According to NASA's Exoplanet Archive notes, all candidates or false positives were excluded from the final planet sample.

\subsection{Uncertainties in the Derived Parameters}
Uncertainties in $T_{\mathrm{eff}}$, $\log g$, and microturbulent velocity ($\xi$) were derived with the \texttt{q$^2$} tool, based on the formalism presented in \cite{epstein2010} and \cite{bensby2014}. Uncertainties in the iron abundances, A(Fe I) and A(Fe II), and A(Mg I) were calculated by combining measurement errors from equivalent width analysis with uncertainties in the stellar parameters. Table \ref{tab:parameters} presents the individual uncertainty components for each parameter.

The typical (median) uncertainties corresponding to the stellar parameters obtained in this analysis are reported in Table \ref{tab:error_budget}: $\delta T_{\rm eff}=39$ K, $\delta \log g = 0.08$ dex, $\delta A(Fe)=0.03$, $\delta \xi = 0.12$ km s$^{-1}$, and $\delta A(Mg) = 0.06$. As in our previous studies \citep{loaiza2023ApJ...946...61L,loaiza2024ApJ...970...53L}, we calculated uncertainties in the derived parameters using the isochrone method, M$_{star}$ and R$_{star}$, by combining the contributions of individual error sources. The contribution resulting from errors in the parallaxes corresponds to a median error of 0.02 mas, representing a 0.46\% error in the stellar mass and radius. 
We adopt a median uncertainty of 0.07 mag in V, which corresponds to the median of the reported V-band errors for our sample, then the associated contribution to the stellar radius uncertainty is approximately 0.54\%.

The contributions of individual error sources to the stellar radius derived using the Stefan–Boltzmann equation depend on $m_K$, $A_K$, $\mu$, $T_{\rm eff}$, and $BC_K$. For our target stars, the typical uncertainty in the K-band magnitude ($m_K$) from 2MASS \citep{skrutskie2006} is 0.021 mag, which translates into an estimated $\sim$ 1 \% contribution to the total error in the stellar radius. The impact of uncertainties in the extinction $A_K$ on the stellar radius is negligible, as $A_K$ values for the target stars are low, ranging from 0.004 to 0.236 mag, with a median of 0.014 mag. 
Neglecting extinction entirely would result in an error of 0.64\%. Uncertainties in the K-band bolometric corrections ($BC_K$) are primarily driven by errors in the effective temperature. To quantify this effect, we varied the $T_{\rm eff}$ of a solar-type star by 39 K—the median uncertainty in our sample—and evaluated the resulting change in $BC_K$, as done by \cite{fulton2018}. We found a typical shift of $\sim$0.02 mag in $BC_K$, corresponding to an uncertainty of approximately 0.98\% in the stellar radius derived. Variations in $\log g$ and [Fe/H] had minimal influence and were considered negligible. The distances and their associated uncertainties for the target stars were adopted from \cite{bailerjones2021}. These correspond to a median uncertainty of 0.005 mag in the distance modulus ($\mu$), which translates into a 0.07\% contribution to the stellar radius uncertainty. The internal precision in effective temperatures in this study is 39 K (median), which contributes approximately 1.4\% to the uncertainty in $R_{\star}$.
The precision of stellar radii and transit depth measurements ($\Delta$F) directly impacts the accuracy of planetary radius estimates. In our sample, the median internal uncertainty in stellar radii is 1.6\%. Transit depth values and their uncertainties were compiled from multiple sources, including \citet{kruse2019}, \citet{guerrero2021ApJS..254...39G}, \citet{livingston2018}, \citet{christiansen2022AJ....163..244C}, \citet{yu2018AJ....156..127Y}, \citet{mayo2018}, \citet{vanderburg2016}, \citet{adams2021PSJ.....2..152A}, \citet{kokori2023ApJS..265....4K}, and \citet{pope2016MNRAS.461.3399P}, among others. Based on this dataset, we estimate a median internal uncertainty of 2.9\% in $\Delta$F. When combined, the uncertainties in stellar radii and transit depths result in typical internal uncertainties in planetary radii ($R_{\rm pl}$) of 2.5\%, 2.6\%, and 7.2\% when using stellar radii derived from SB, \texttt{PARAM}, and \texttt{isochrones}, respectively. A summary of the error contributions to both $R_\star$ and $R_{\rm pl}$ is provided in Table \ref{tab:error_budget}.
\begin{deluxetable}{ll}
\tablecaption{Error budget\label{tab:error_budget}}
\tablecolumns{2}
\tablenum{5}
\tablewidth{0pt}
\tablehead{
\colhead{Parameter} & \colhead{Median Uncertainty}
}
\startdata
$T_{\rm eff}$  & 39 K \\
$\log g$ & 0.08 dex \\
A(Fe)         & 0.03 dex \\
A(Mg)         & 0.02 dex \\ 
$V$             & 0.07 mag \\
plx           & 0.02 mas \\
$m_K$         & 0.021 mag \\
$A_K$         & 0.014 mag \\
$BC_K$        & 0.02 mag \\
$\mu$         & 0.005 mag \\
M$_{star}$    &   \texttt{PARAM} = 0.02 M$_\odot$ / \texttt{isochrones} = 0.03 M$_\odot$ \\
$\Delta$F     & 2.88 \% \\
R$_{star}$     & SB = 1.59 \% / \texttt{PARAM} = 1.97 \% / \texttt{isochrones} = 7.20\% \\
R$_{pl}$      & SB = 2.48 \% / \texttt{PARAM} = 2.57 \% / \texttt{isochrones} = 6.65 \% \\
\enddata
\end{deluxetable}

\begin{deluxetable}{lrccc}
\tabletypesize{\scriptsize}  
\setlength{\tabcolsep}{2pt}  
\tablecaption{Planetary Radii\label{tab:planet_rad}}
\tablecolumns{5}
\tablenum{6}
\tablewidth{0pt}
\tablehead{
\colhead{} & \colhead{} & \colhead{SB} & \colhead{\texttt{PARAM}} & \colhead{\texttt{isochrones}} \\
\colhead{Planet Name} & $\Delta F$ (ppm)&   \colhead{$R_{pl}$ ($R_\oplus$)} & \colhead{$R_{pl}$ ($R_\oplus$)} & \colhead{$R_{pl}$ ($R_\oplus$)} \\
}
\startdata
K2-44 b & 334.2 & $3.028 \pm 0.067$ & $2.968 \pm 0.079$ & 3.349$^{+0.280}_{-0.327}$ \\
K2-5 b & 1496.0 & $2.150 \pm 0.059$ & $1.664 \pm 0.047$ & 2.245$^{+0.096}_{-0.100}$ \\
K2-5 c & 1378.0 & $2.063 \pm 0.065$ & $1.597 \pm 0.054$ & 2.154$^{+0.098}_{-0.104}$ \\
K2-6 b & 724.0 & $2.612 \pm 0.054$ & $2.606 \pm 0.076$ & 2.991$^{+0.311}_{-0.494}$ \\
K2-7 b & 681.0 & $4.489 \pm 0.173$ & $3.915 \pm 0.117$ & 4.637$^{+0.415}_{-0.792}$ \\
\nodata  & \nodata  & \nodata & \nodata  & \nodata  \\    
\enddata
\tablecomments{Transit depth ($\Delta F$) collected from \cite{kruse2019}, \cite{guerrero2021ApJS..254...39G}, \cite{livingston2018}, \cite{christiansen2022AJ....163..244C}, \cite{yu2018AJ....156..127Y}, \cite{mayo2018}, \cite{vanderburg2016}, among others. This table is published in its entirety in machine-readable format. A portion is shown here for guidance regarding its form and content.}
\end{deluxetable}

\section{Computation of activity indices} \label{sec:S_ind}
The chromospheric indicators utilized to characterize stellar activity in our sample include the Ca II H \& K and H$\alpha$ lines. For each of these spectral features, a detailed methodology is provided, describing the process by which the corresponding activity index was derived.

\subsection{The Ca II H\&K Index} \label{sec:calcio}
The Ca II H \& K S-index, as introduced by the Mount Wilson Observatory program, is among the most widely used diagnostics of chromospheric activity. The S-index is defined as a ratio of fluxes measured in the H and K line cores and fluxes in nearby pseudo-continuum passbands, which are labeled R and V,

\begin{equation}\label{eq:s_inst}
    S_{HK} = \frac{H+K}{R+V}.
\end{equation}

Following the methodology of \citet{isaacson2010ApJ...725..875I}, we measured the instrumental S-index from normalized spectra using Equation (\ref{eq:s_inst}). In this procedure, the fluxes $H$ and $K$ are obtained from the line cores using triangular bandpasses (FWHM = 1.09 \AA) centered at $\lambda=3968.47$ \text{\AA} and $\lambda=3933.67$\AA\, respectively. The pseudo-continuum fluxes $R$ and $V$ are measured in 20 \AA --wide rectangular bandpasses centered at $\lambda=3901$ \text{\AA} and $\lambda=4001$ \AA, respectively. In Equation (\ref{eq:shk}), the coefficients $C_2$ and $C_3$ are then determined from the ratios of the line-core fluxes to the continuum fluxes.

\begin{equation}\label{eq:shk}
    \rm S_{HK} = C_1 \frac{(H + C_2K)}{(R + C_3V)} + C_4.
\end{equation}

We estimated the coefficients $C_1$ and $C_4$ by minimizing the differences between the values from the Mt. Wilson survey \citep{duncan1991ApJS...76..383D} and the $S_{HK}$ determined from the HIRES spectra of 36 stars observed by both, using Equation (\ref{eq:shk}). This was done using the Markov Chain Monte Carlo (MCMC) technique using \texttt{emcee} package \citep{foreman2013PASP..125..306F}. This procedure was also performed by \cite{mayo2018} to calibrate the TRES and Mt. Wilson $S_{HK}$ data. 
Finally, the calibrated relationship using the MCMC parameters is given by (\ref{eq:shk_cal}). 

\begin{equation}\label{eq:shk_cal}
    \rm S_{HK} = 11.382 \frac{(H + 0.331K)}{(R + 1.156V)} + 0.092
\end{equation}

To verify our calibration, we used the star Tau Ceti. This is one of the most extensively studied stars in investigations of stellar activity and magnetic cycles, with long-term data on the $S_{HK}$ index. It is also a low-activity star, which enables us to assess the accuracy of the $S_{HK}$ index calibration. The mean value of the $S_{HK}$ index found for Tau Ceti, based on 19 spectra obtained over six nights of observations spanning six years, is $S_{HK} = 0.178$, and its mean absolute deviation is 
0.017. Our result is similar to that found by \cite{gomes2021AA...646A..77G}, $S_{HK} = 0.175 \pm 0.0015$ (mean and standard deviation), and \cite{isaacson2024ApJ...961...85I}, $S_{HK} = 0.172 \pm 0.0023$ (median and standard deviation). We note that the differences between the standard deviations reported in those studies and our MAD are due to the observational time span, as \citet{gomes2021AA...646A..77G} used a 14-year baseline (595 observing nights) and \citet{isaacson2024ApJ...961...85I} used 15 years of data (942 observing nights).

For stars with TRES spectra, we used the calibration from \cite{mayo2018} to determine the Ca II index, which is defined as:
\begin{equation}
    \rm S_{HK}(TRES) = 15.496 \frac{(H + 0.876K)}{(R + 0.775V)} - 0.0031.
\end{equation}

We measured the $\rm S_{HK}$ index in a total of 2,385 spectra from 301 host stars.

The S-index provides a flux that contains contributions from both the chromospheric and photospheric components, which are referred to as $R^\prime_{\rm HK}$ and $R_{\rm phot}$, respectively. To isolate the chromospheric signal for stars with different effective temperatures, it is necessary to subtract the photospheric contribution by accounting for the temperature dependence of the R and V fluxes. We adopt the method described by \citet{noyes1984ApJ...279..763N} and, following \citet{loaiza2024ApJ...970...53L}, apply the bolometric corrections proposed by \citet{rutten1984AA...130..353R}. These corrections are defined for $0.3 \leq (B-V) \leq 1.7$ and apply to dwarf and giant stars. The resulting mean $\log R^\prime_{\rm HK}$ values for our sample are presented in Table \ref{tab:parameters}.

\subsection{The H$\alpha$ Index} \label{sec:halpha}
The H$\alpha$ index is a well-established indicator of chromospheric activity \citep{linsky2017ARAA..55..159L}. In this work, we computed the H$\alpha$ index following the method of \cite{boisse2009AA...495..959B}, but with a slightly broader central band because it better captures the chromospheric activity contribution \citep{gomes2011AA...534A..30G}. The index is defined as: 

\begin{equation}
   H\alpha  = \frac{F_{H\alpha}}{F_1 + F_2},
\end{equation}

\noindent where $F_{H\alpha}$ is the flux on the H$\alpha$ line, and $F_1$ and $F_2$ are the continuum fluxes on either side of the line. The $F_{H\alpha}$ interval is 1.6 \AA\ wide, centered at 6562.808 \AA, while the $F_1$ interval is 10.75 \AA\ wide, centered at 6550.87 \AA, and the $F_2$ interval is 8.75 \AA\ wide, centered at 6580.309 \AA. 
Because continuum flux varies with stellar color, systematic differences arise in the measured mean H$\alpha$ level, making direct comparisons across stars difficult \citep[e.g.,][]{cinecungui2007AA...469..309C}. It is necessary to account for the photospheric contribution to allow a more meaningful comparison of the H$\alpha$ indices between stars. We therefore applied the correction to the $I_{\text{H}\alpha}$ activity index following Equation~(A.1) from \citet{gomes2014AA...566A..66G}. The resulting H$\alpha$ and $\log I_{\text{H}\alpha}$ values are presented in Table~\ref{tab:parameters}.

\section{Discussion} \label{sec:discus}

\subsection{Stellar Parameters, Radii and Masses: Comparisons with the Literature}

We determined stellar radii using both the Stefan-Boltzmann law and an isochrone fitting approach implemented in various Bayesian codes such as \texttt{PARAM} and \texttt{isochrones}. The \texttt{isochrones} package is based on MIST stellar isochrones, and \texttt{PARAM} uses MESA isochrones. 
The two codes take as input the spectroscopic parameters ($T_{\rm eff}$ and [Fe/H]), parallaxes from Gaia DR3, and $V$ magnitudes.

\begin{figure}
    \centering
    \includegraphics[width=0.82\linewidth]{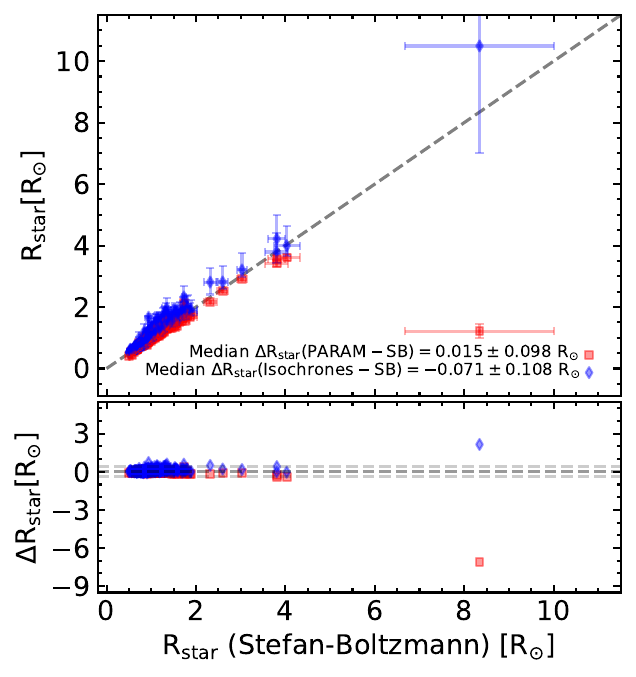}
    
    \includegraphics[width=0.80\linewidth]{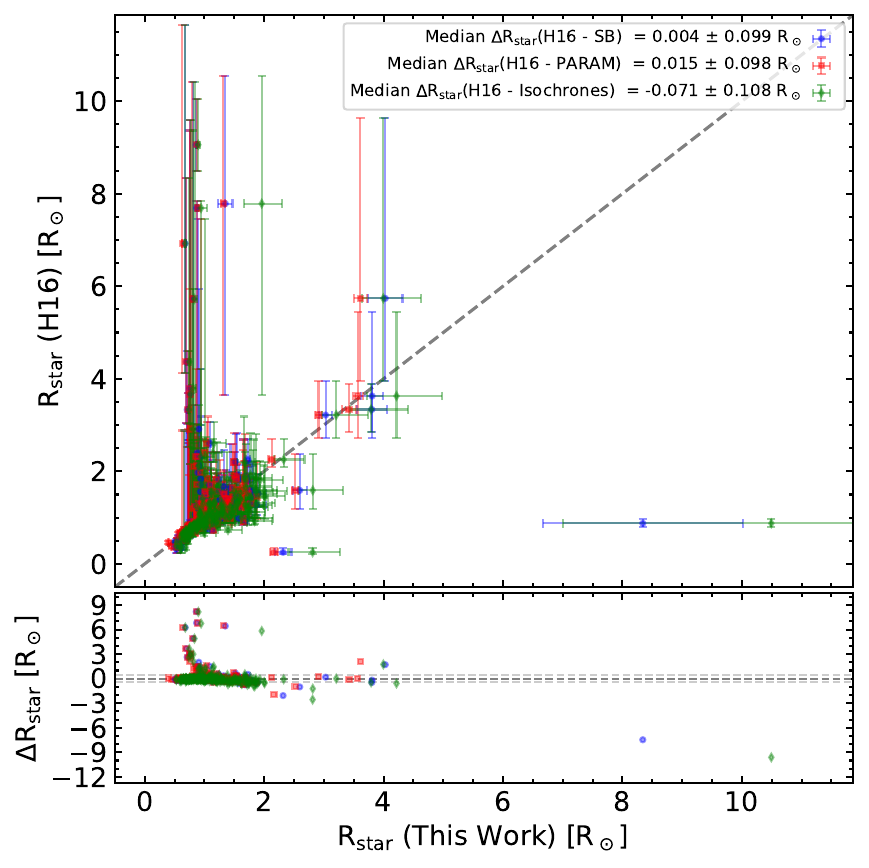}
    
    \includegraphics[width=0.80\linewidth]{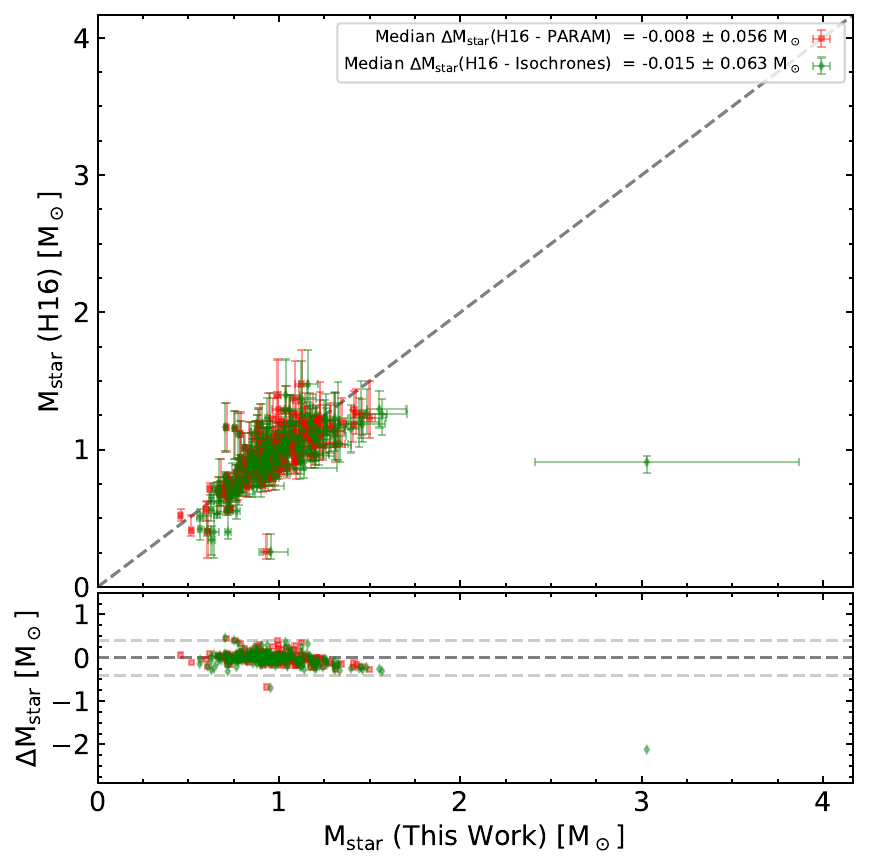}
    \caption{Comparison between the stellar radii determined in this work using the Stefan-Boltzmann law (SB) and using the isochrone method implemented in \texttt{PARAM} and \texttt{isochrones} (left panel). Comparisons of the stellar radii and masses in this study with those determined through asteroseismology by \citep[][H16]{huber2016ApJS..224....2H} (middle and right panels).}
    \label{fig:R_M_H16}
\end{figure}

Figure \ref{fig:R_M_H16} (top panel) shows the differences among the various stellar radius estimates as a function of the Stefan-Boltzmann derived radius.  
The median difference and median absolute deviation (MAD) between the stellar radius derived with the isochrone method and the Stefan-Boltzmann equation (\texttt{PARAM}: $0.015\pm 0.098$ R$_\odot$ and \texttt{isochrones}: $-0.071\pm 0.108$ R$_\odot$) indicate that stellar radii determined with \texttt{PARAM} tend to be slightly smaller than those obtained via SB, while the results using \texttt{isochrones} tend to be higher than those obtained via SB. 
For most stars, the differences remain within the dashed lines corresponding to the $2\sigma$ confidence level (considering that the median of the errors using and $10.493^{+4.658}_{-3.490}$ R$_\odot$ (\texttt{isochrones}). We can notice that the uncertainties associated with the measurements with SB and \texttt{isochrones} are larger than the uncertainty with \texttt{PARAM}. We used the same spectroscopic parameters for the three cases. However, for SB, we employed the distance, whereas for the isochrones method, we used the parallax. From Gaia DR3 this star has a parallax of $0.073593 \pm 0.140761$ mas, and according to \cite{bailerjones2021}, this star has a distance of 4573.30$_{-478.42}^{+473.20}$ pc. This star also was analyzed in \cite{soto2018MNRAS.478.5356S}, in which they report a distance of $453^{+72}_{-46}$ pc. Using this distance, the Stefan-Boltzmann method produced a significantly smaller radius of 0.83 R$_\odot$. This comparison suggests that the large radius values obtained using the Bailer-Jones distance and Gaia DR3 parallax are likely overestimated. On the other hand, according to \citep{evans2018RNAAS...2...20E}, this star has a known companion at $0.35''$, which is consistent with the high RUWE value of 4.818 reported in Gaia DR3. Here, RUWE (re-normalized unit weight error) is a parameter used to explore stellar multiplicity \citep[e.g.,][]{castro_ginard2024A&A...688A...1C}. This elevated RUWE may indicate unresolved stellar multiplicity, which could systematically affect the parallax measurement and contribute to the observed discrepancies in the derived stellar radius.

In the middle panel of Figure \ref{fig:R_M_H16}, we compare our stellar radii (SB, \texttt{PARAM}, and \texttt{isochrones}) with those determined by asteroseismology \citep[][H16]{huber2016ApJS..224....2H}. The comparison yields the following median differences and median absolute deviations (MAD): 0.004 $\pm$ 0.099 R$_\odot$, 0.015 $\pm$ 0.098 R$_\odot$, and -0.071 $\pm$ 0.108 R$_\odot$, respectively for SB, \texttt{PARAM} and \texttt{isochrones}. 
Most of the results are consistent within 2$\sigma$ (where $\sigma = 0.2$ R$_\odot$, the mean stellar radius obtained by asteroseismology), but there are a few outliers. 
The star K2-238 is again identified. 
Using the distance from \citet{soto2018MNRAS.478.5356S}, its stellar radius is estimated to be 0.83 R$_\odot$, which is in very good agreement with the value from asteroseismology of 0.873 R$_\odot$ reported by \citet{huber2016ApJS..224....2H}. 

Another star with a large stellar radius determined here is K2-171; it has a parallax of $\pi = 1.666$ mas, corresponding to a distance of $d = 593$ pc, an apparent magnitude $V = 12.745$, and an absolute magnitude $M_V = 3.76$. These parameters indicate that the star is more luminous and larger than a typical K-dwarf. According to results reported by Gaia DR2, \cite{mayo2018}, and \cite{kruse2019}, the stellar radius is estimated as $2.378^{+0.038}_{-0.018}$ R$_\odot$, $1.717^{+0.221}_{-0.194}$ R$_\odot$, and $2.378^{+0.018}_{-0.038}$ R$_\odot$, respectively. Such a radius is significantly larger than the 0.250 R$_\odot$ as reported by \cite{huber2016ApJS..224....2H}, supporting the conclusion that K2-171 is more likely a K-type subgiant rather than a K-dwarf.

K2-99 is also in the group having a large estimated radius; it has a parallax of $\pi = 1.932$ mas, corresponding to a distance of $d = 509$ pc, an apparent magnitude $V = 11.149$, and an absolute magnitude $M_V = 2.53$ mag. These values suggest that the star is significantly more luminous than a main-sequence dwarf and is instead consistent with a subgiant or slightly evolved star. The stellar radius for K2-99 has been independently estimated by several works: Gaia DR2 ($2.709^{+0.134}_{-0.415}$ R$_\odot$), \citet[][1.98 R$_\odot$]{petigura2018}, \citet[][$2.63 \pm 0.07$ R$_\odot$]{livingston2018}, \citet[][$2.276^{+0.320}_{-0.257}$ R$_\odot$]{mayo2018}, \citet[][$2.71^{+0.42}_{-0.13}$ R$_\odot$]{kruse2019}, and \citet[][$2.55 \pm 0.02$ R$_\odot$]{smith2022MNRAS.510.5035S}. These values are all significantly higher than the estimate from \cite{huber2016ApJS..224....2H}, who reported a radii of $1.589^{+0.781}_{-0.395}$ R$_\odot$. The more recent and precise determinations strongly support a larger stellar radius, providing support for our result of $2.597\pm0.116$, $2.519\pm0.051$, and $2.814598^{+0.512}_{-0.223}$ for K2-99.

With a parallax of $\pi = 5.619$ mas, K2-37 is at a distance of $d = 177$ pc and has an absolute magnitude of $M_V = 5.48$ mag. \cite{huber2016ApJS..224....2H} reported a surface gravity of $\log g = 2.464 \pm 0.048$, whereas in our study we determined a significantly higher value of $\log g = 4.52 \pm 0.017$. This value is in agreement with the results of \cite{mayo2018}, \cite{crossfield2016}, and \cite{sinukoff2016}, all of whom analyzed high-resolution spectra. The stellar radii reported in Gaia DR2 ($0.792,R_\odot$), \citet[][$0.818,R_\odot$]{mayo2018}, \citet[][$0.870,R_\odot$]{crossfield2016}, \citet[][$0.85,R_\odot$]{sinukoff2016}, and \citet[][$0.792,R_\odot$]{kruse2019} are all consistent with our results.

Other stars with the larger stellar radii determined in \cite{huber2016ApJS..224....2H} are those stars with the greatest uncertainties: K2-216 ($\pi=8.688$ mas, $d=114$ pc, $M_V=7.04$ mag), K2-58 (5.493 mas, 181 pc, 5.98 mag), K2-275 (8.063 mas, 123 pc, 6.52 mag), K2-203 (5.915 mas, 168 pc, 6.14 mag), K2-268 (3.073 mas, 321 pc, 6.22 mag), K2-395 (3.451 mas, 286 pc, 6.05 mag), K2-105 (5.011 mas, 198 pc, 5.19 mag), and K2-201 (5.093 mas, 195 pc, 5.34 mag). According to their parallaxes, distances, and absolute magnitudes, these stars are placed squarely in the realm of G- and K-dwarfs, where a subsolar radius is expected. This provides confidence in our derived radii and aligns more closely with our $\log g$ values. 

For stellar masses (bottom panel of Figure 2), the median differences between the values determined via asteroseismology \citep{huber2016ApJS..224....2H} and those obtained with \texttt{PARAM} and \texttt{isochrones} are $-0.008 \pm0.056$ M$_\odot$ and $-0.015\pm0.063$ M$_\odot$, respectively. As was the case in the radius comparison, K2-171 again appears as an outlier when comparing the stellar masses derived in H16 with our results from \texttt{PARAM} and \texttt{isochrones}, suggesting that this star is more massive than reported by H16. The stellar masses we determine for this star are 0.931 M$_\odot$ with \texttt{PARAM} and 0.954 M$_\odot$ with \texttt{isochrones}, consistent with the value reported by \citet[][0.892$^{+0.060}_{-0.027}$ M$_\odot$]{mayo2018}. The other outlier in the comparison between H16 and \texttt{isochrones} is K2-238. As previously mentioned in the radius comparison, the parallax of this star is not reliable. Overall, our stellar masses obtained with \texttt{PARAM} and \texttt{isochrones} are consistent with the asteroseismic values within the uncertainties.

In this study, we have 52 stars in common with the recent work of \cite{howard2025ApJS..278...52H}, who analyzed Keck HIRES spectra and derived stellar parameters from an independent analysis based on the spectral synthesis method, and using the codes \texttt{Spectroscopy Made Easy} \citep{piskunov2017}, \texttt{SpecMatch} \citep{petigura2015PhDT........82P} and \texttt{SpecMatch-Emp}  \citep{yee2017ApJ...836...77Y}. 
A comparison between the results of this work and \cite{howard2025ApJS..278...52H} reveals good agreement within the reported uncertainties, indicating that the two different methodologies adopted in the analyses of high-resolution spectra produce consistent results.
The stellar parameters show median offsets (Howard et al. - This Work) of $\Delta T_{\rm eff} = -35 \pm 45$ K, $\Delta \log g = 0.080 \pm 0.100$ dex, and $\Delta[\mathrm{Fe/H}] = -0.030 \pm 0.043$ dex. Stellar radii differ by $\Delta R_\star = 0.007 \pm 0.015$ R$_\odot$, $-0.003 \pm 0.044$ R$_\odot$, and $-0.106 \pm 0.070$ R$_\odot$ for the SB, \texttt{PARAM}, and \texttt{isochrone} methods, respectively. Similarly, stellar masses differ by $\Delta M\star = -0.029 \pm 0.027$ M$_\odot$ and $-0.028 \pm 0.019$ M$_\odot$ for \texttt{PARAM} and \texttt{isochrones}, respectively. For planetary radii, we obtain median differences of $-0.212 \pm 0.169$ R$_\oplus$, $-0.233 \pm 0.204$ R$_\oplus$, and $-0.650 \pm 0.335$ R$_\oplus$, for SB, \texttt{PARAM}, and \texttt{isochrone} methods, respectively. Finally, for the stellar activity indices, we find median differences of $\Delta S_{\rm HK} = -0.016 \pm 0.012$ and $\Delta \log R^{\prime}_{\rm HK} = -0.073 \pm 0.075$.

We also have 144 stars in common with the K2 study of \cite{zink2023AJ....165..262Z}, who  determined stellar parameters using the spectral synthesis method and the \texttt{SpecMatch-synthetic} code. A comparison of the results for stellar parameters exhibits small differences with median offsets (Zink et al. - This Work) of $\Delta T_{\rm eff} = -15 \pm 66$ K, $\Delta \log g = 0.100 \pm 0.140$ dex, $\Delta \mathrm{[Fe/H]} = 0.029 \pm 0.050$ dex, $\Delta R_\star$ = $0.004 \pm 0.018$ R$_\odot$ (SB), $0.025 \pm 0.033$ R$_\odot$ (\texttt{PARAM}), $-0.066 \pm 0.050$ R$_\odot$ (\texttt{isochrones}), and $\Delta M_\star$ = $-0.009 \pm 0.036$ M$_\odot$ (\texttt{PARAM}), $-0.011\pm0.034$ M$_\odot$ (\texttt{isochrones}), which overall indicate good agreement between the two studies.

\subsection{The Exoplanetary Radius gap}
\begin{figure}[!h]
    \centering
    \includegraphics[width=\linewidth]{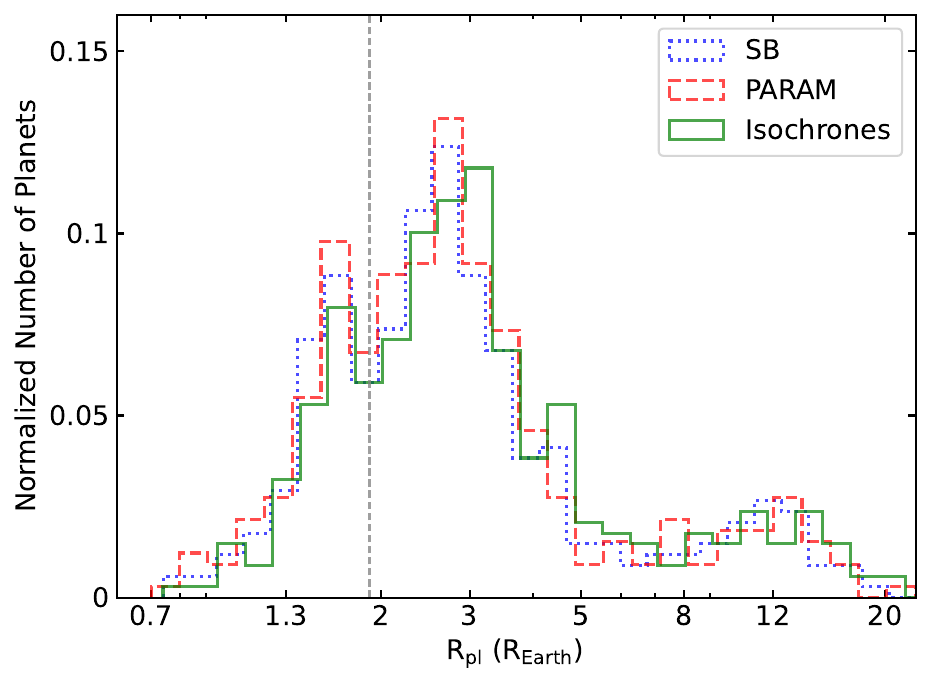}
    \caption{Distribution of planetary radii of our exoplanets sample, derived using stellar radii estimated from the Stefan-Boltzmann equation and the isochrones method using \texttt{PARAM} and \texttt{isochrones} packages. The vertical line highlights the radius gap region.} 
    \label{fig:Rpl_dist}
\end{figure}

Figure \ref{fig:Rpl_dist} shows three normalized histograms representing the distribution of planetary radii ($R_{pl}$) obtained using different stellar radius estimation methods. The blue histogram corresponds to 339 planets, based on stellar radii for 257 stars, as determined via the SB equation. The red histogram shows the distribution for 327 planets obtained using \texttt{PARAM} radii for 247 stars, while the green histogram displays the results for 339 planets using isochrones-based radii for 257 stars. Similarly to what was found in our previous study of K2 targets \citep{loaiza2023ApJ...946...61L,loaiza2024ApJ...970...53L} with a sample of less than 100 exoplanets, we identify the radius gap in $\sim 1.9$ R$_\oplus$ in all distributions. 

This analysis was carried out by varying only the stellar radii since the value of transit depth remained unchanged for all stellar radius estimates. Therefore, the differences in the peaks and valleys among the distributions arise solely from the choice of stellar radius. All three distributions exhibit the radius gap within the expected range and show the best agreement with the established location of the gap \citep{fulton2017,owen2017,vaneylen2018,Martinez2019}. 

\subsection{Stellar abundances}

In the the first row panel of Figure \ref{fig:cdf_feh}, we show the metallicity distributions for three different samples: APOGEE red giants (black), Kepler CKS stars \citep[red;][]{Ghezzi2021}, and our sample of K2 stars (blue, this work). 
The metallicity distribution of our K2 sample covers the range between [Fe/H] = -1.06 to 0.58 (iron abundances roughly between $6.50 <A(Fe) < 8.00$), with $\sim$21\% of stars being more metal-poor than [Fe/H] = -0.2 dex.
The median ($\pm$MAD) of the metallicity of the distribution is [Fe/H] = 0.004 $\pm$ 0.164 dex (A(Fe) = $7.526 \pm 0.164$ dex), and this value is very close to the metallicity obtained here for solar-proxy asteroids (A(Fe) = 7.52), and also in good agreement with the solar Fe abundance of \citep{magg2022AA...661A.140M}. The metallicity range of our K2 sample largely overlaps that of the Galactic thin disk, though it is slightly more metal-poor. 

The comparison of the metallicity distributions of the K2 sample with that of the CKS sample from \citet{Ghezzi2021} reveals that the K2 stars in our study tend to be more metal-poor; this was also noted in \citet{loaiza2023ApJ...946...61L}.
This result is consistent with the conclusion from \citet{Ghezzi2021} that the metallicity distribution of the CKS sample closely resembles the metallicity distribution function (MDF) of stars in the solar neighborhood, as determined by stellar populations located at Galactocentric distances between 7 and 9 kpc in the APOGEE and GALAH surveys \citep{hayden2015ApJ...808..132H,hayden2020MNRAS.493.2952H}. 
For reference, the MDF of the APOGEE red giants is shown as the black curve in Figure \ref{fig:cdf_feh}. We can see that the distributions for the three cases have peaks at 0.00 dex for K2, 0.05 dex for APOGEE, and 0.10 dex for \citet{Ghezzi2021}. 

Figure \ref{fig:Mg_FeH} (left panel) illustrates the distribution of the $\alpha$-element magnesium as a function of [Fe/H], with the stellar sample analyzed in this work shown in blue and the APOGEE DR17 magnesium abundance data \citep{abdurrouf2022ApJS..259...35A} shown in gray as background. Our measurements generally follow the well-established abundance patterns observed in the APOGEE survey, demonstrating the good agreement between our spectroscopic analysis and this large-scale stellar near-infrared dataset. The distribution reflects the well-known bimodal structure of the Galactic disk  \citep[e.g.,][]{fuhrmann1998A&A...338..161F,Reddy2006,nidever2014ApJ...796...38N,anders2014AA...564A.115A,Hayes2018ApJ...852...49H,queiroz2020A&A...638A..76Q}. We separate our sample into the high- and low-[Mg/Fe] sequences (generally representing the chemical thick and thin disks) following the criterion of \cite{weinberg2022ApJS..260...32W}. This yields 59 stars in the high-alpha sequence  and 198 stars in the low-alpha sequence. 
The majority of our K2 targets are in the metallicity range between $\rm -0.5 \lesssim  [Fe/H] \lesssim +0.4$ and exhibit [Mg/Fe] abundances typical of the thin disk. 
This result is consistent with the expected chemical properties of planet-hosting stars in the Galactic disk \citep{adibekyan2012A&A...543A..89A,bashi2022MNRAS.510.3449B}. 

In the right panel of Figure \ref{fig:Mg_FeH}, we show [Mg/Fe] versus [Fe/H] for Single systems, where stellar hosts of Small exoplanets are marked in purple and those of Large exoplanets in green. There are 53 host stars in the high-alpha (thick disk) population and 144 host stars in the low-alpha (thin disk) population. In the low-alpha sequence, the proportion of stars hosting Single Small planets relative to those hosting Single Large planets is 3.6, while the corresponding proportion in the high-alpha sequence is 1.5. These results corroborate previous studies on exoplanet occurrence rates, which show that SE and SN are approximately 50\% less common in the thick-disk regime compared to the thin disk \citep{bashi2022MNRAS.510.3449B,zink2023AJ....165..262Z, hallat2025ApJ...979..120H}.

\subsubsection{CDF Comparisons}
\begin{figure*}
    \centering
    \includegraphics[width=0.4\linewidth]{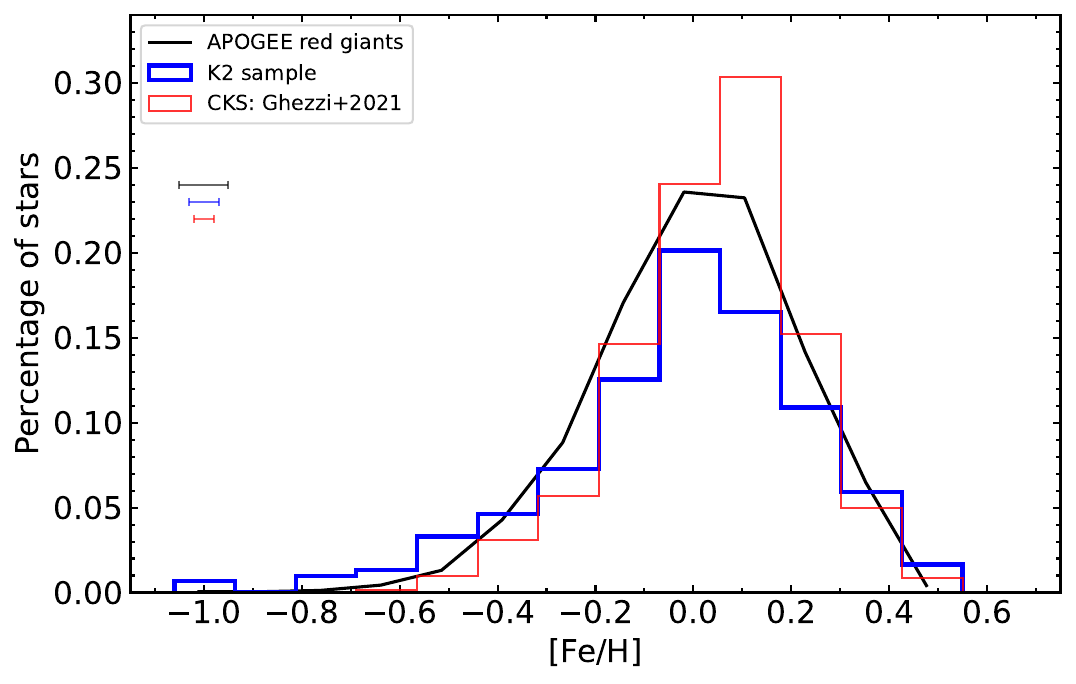}
    
    \includegraphics[width=0.32\linewidth]{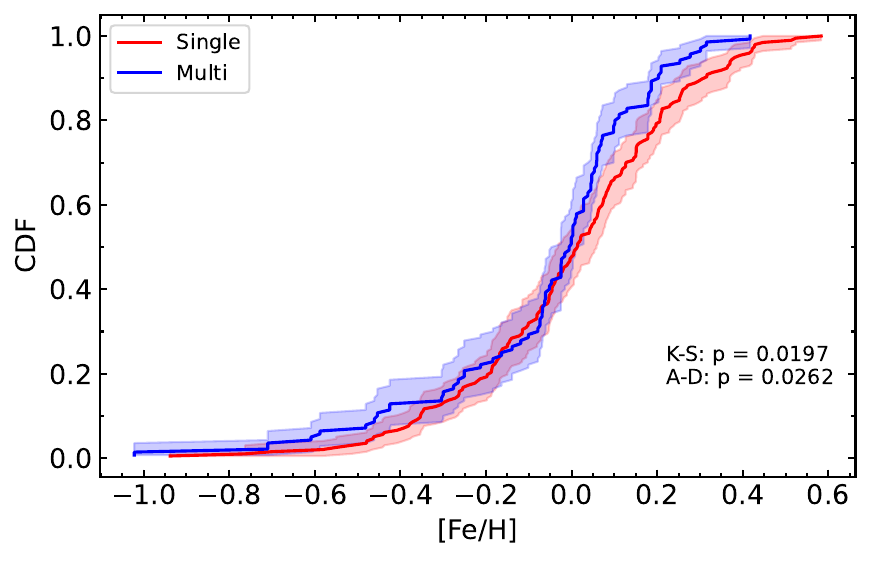}
    \includegraphics[width=0.32\linewidth]{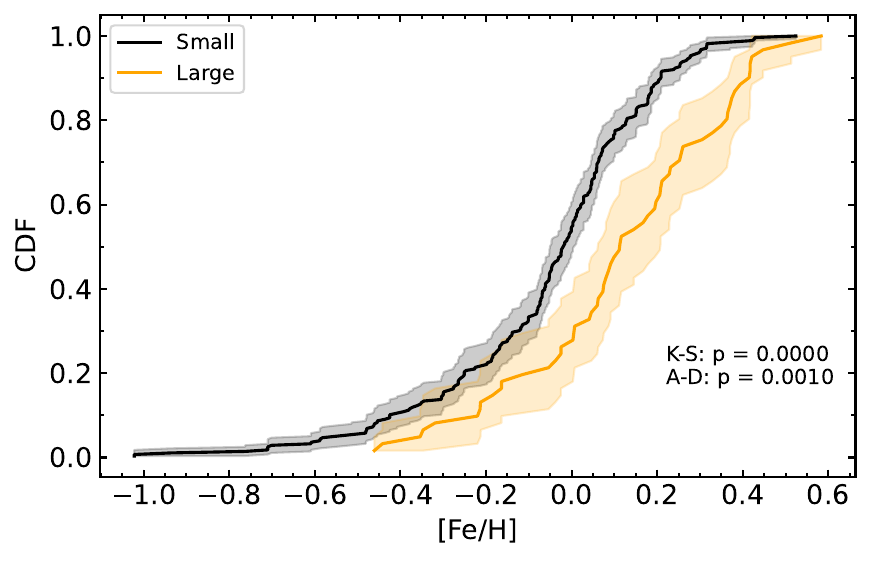}
    \includegraphics[width=0.32\linewidth]{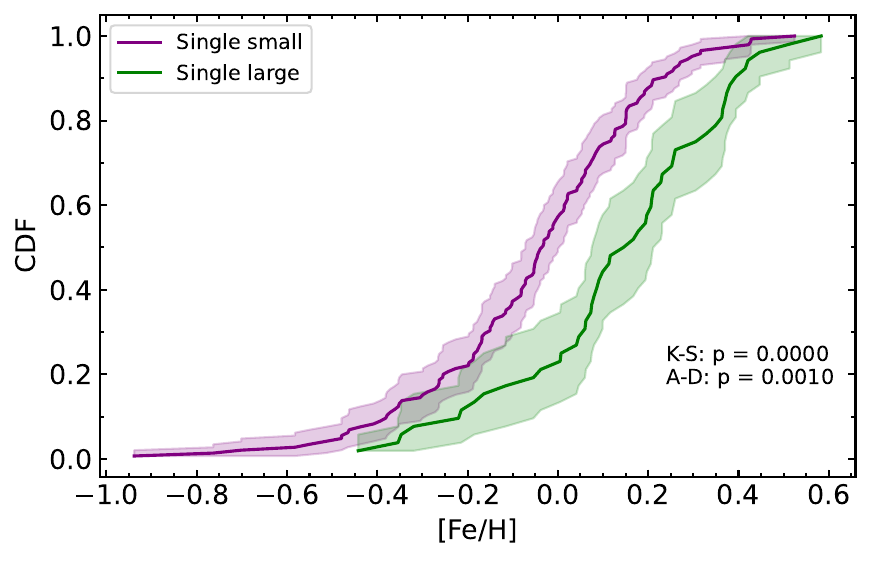}
    \includegraphics[width=0.32\linewidth]{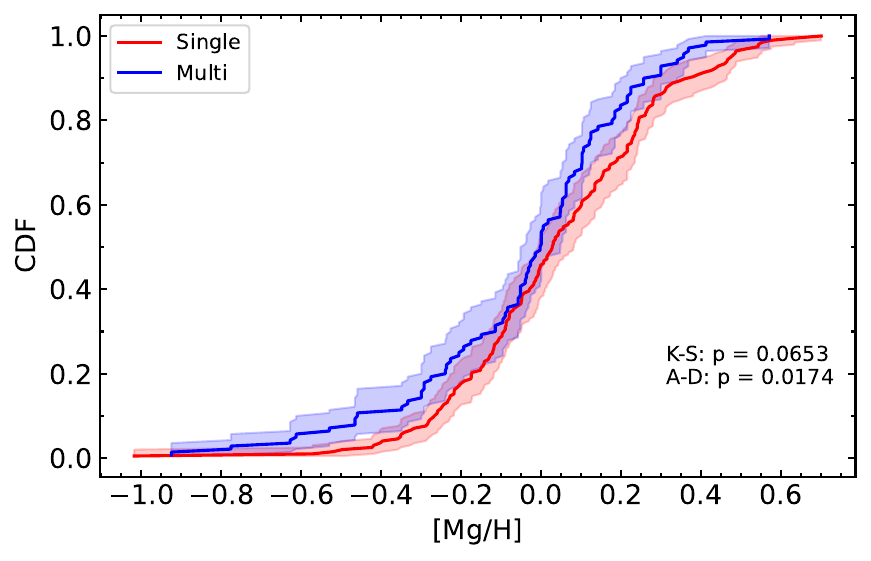}
    \includegraphics[width=0.32\linewidth]{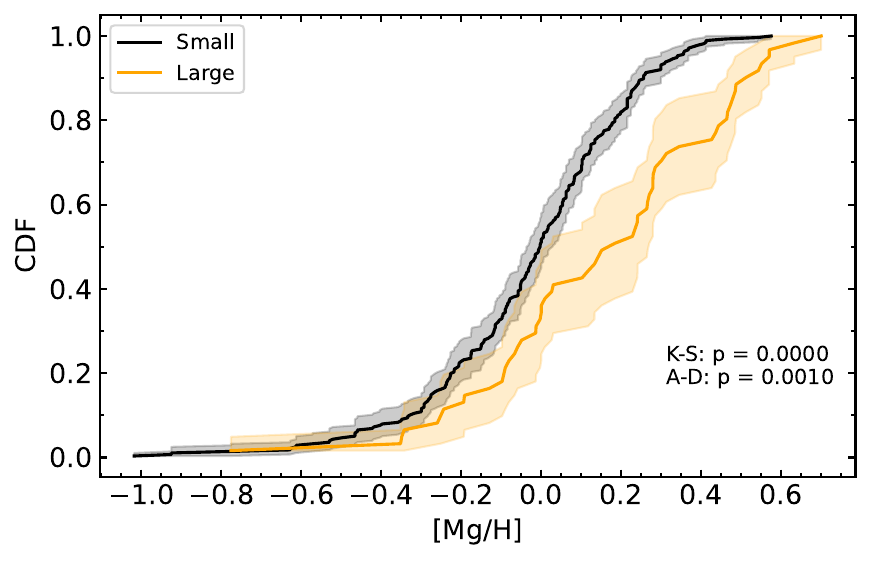}
    \includegraphics[width=0.32\linewidth]{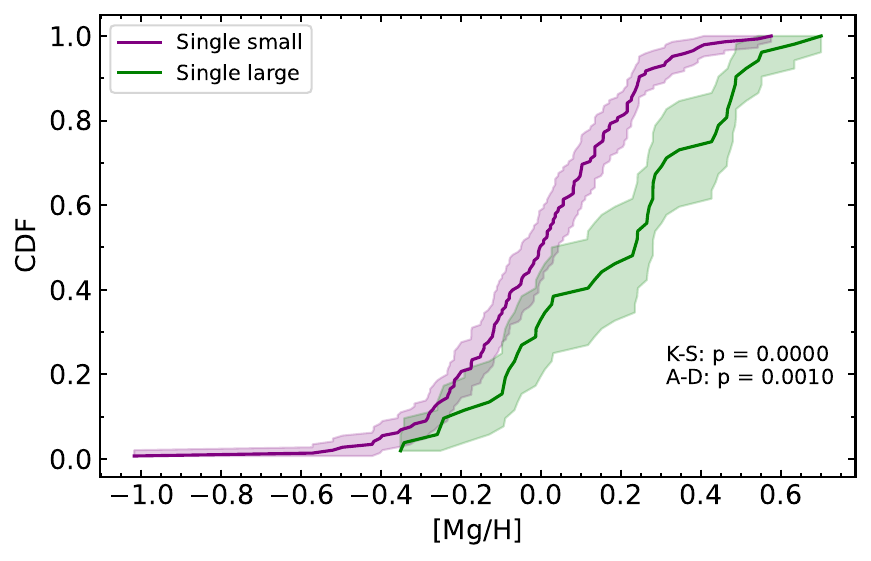}    
    \includegraphics[width=0.32\linewidth]{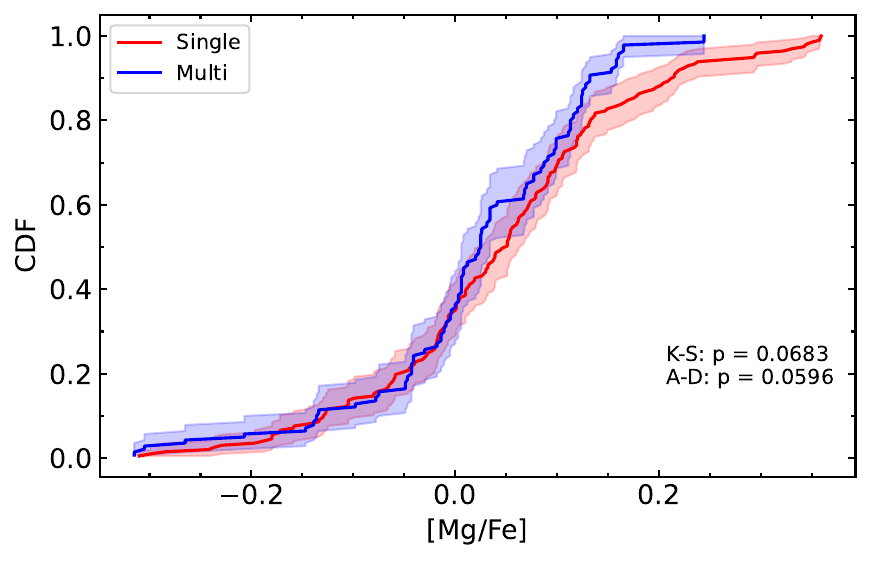}
    \includegraphics[width=0.32\linewidth]{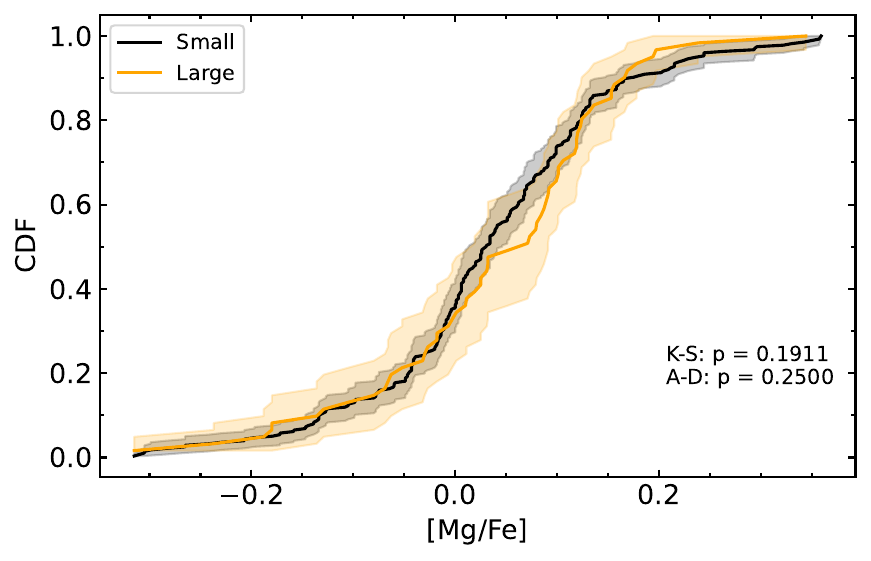}
    \includegraphics[width=0.32\linewidth]{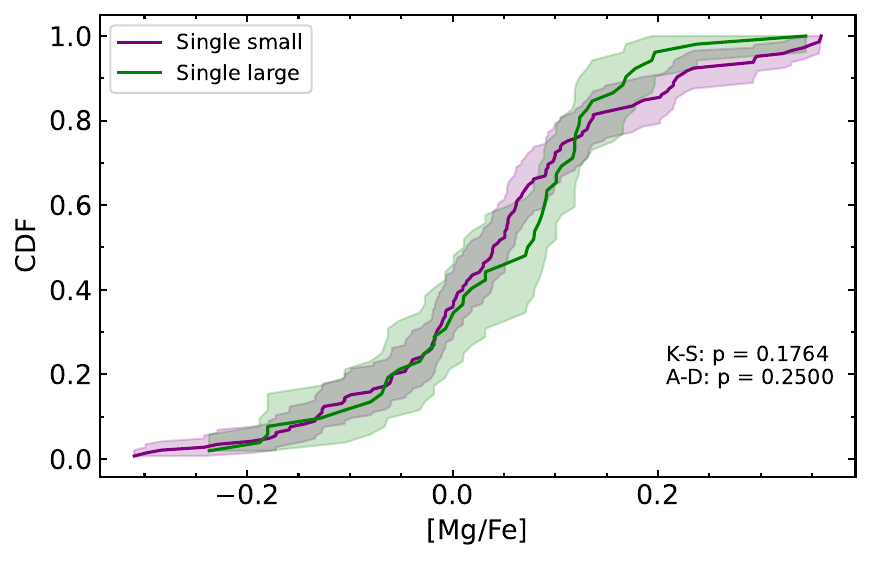}
    \caption{Iron abundance distributions (top left panel) are shown for the K2 host star sample from this study in comparison with the metallicity distributions for the CKS sample from \cite{Ghezzi2021} and red-giants from the APOGEE survey. The median uncertainties in the iron abundances for each sample are shown on the left. In the second row panel, we show the cumulative distribution functions (CDFs) of metallicities of host stars of single planets (red) versus multiple planets (blue), Small planets (black) versus and Large planets (orange), and Single Small (purple) versus Single Large (green). In the third and bottom panels, we show the same cases as for metallicity, but now for [Mg/H] and [Mg/Fe], respectively.  The [Mg/H] distribution is noticeably shifted toward higher values for stars hosting Large planets, but this is not shown for [Mg/Fe]. The shaded regions around each CDF correspond to the 95\% confidence intervals calculated via bootstrap.}
    \label{fig:cdf_feh}
\end{figure*}

\begin{figure*}
    \centering
    \includegraphics[width=0.45\linewidth]{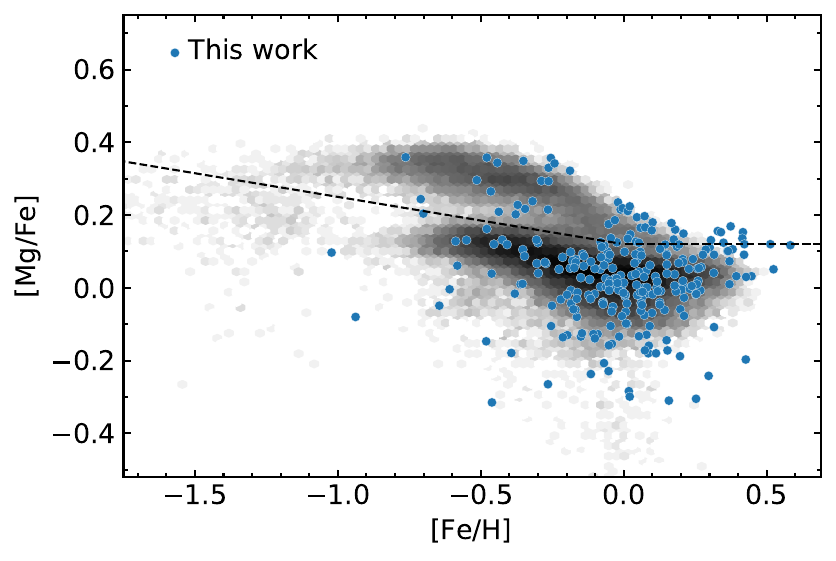}
    \includegraphics[width=0.45\linewidth]{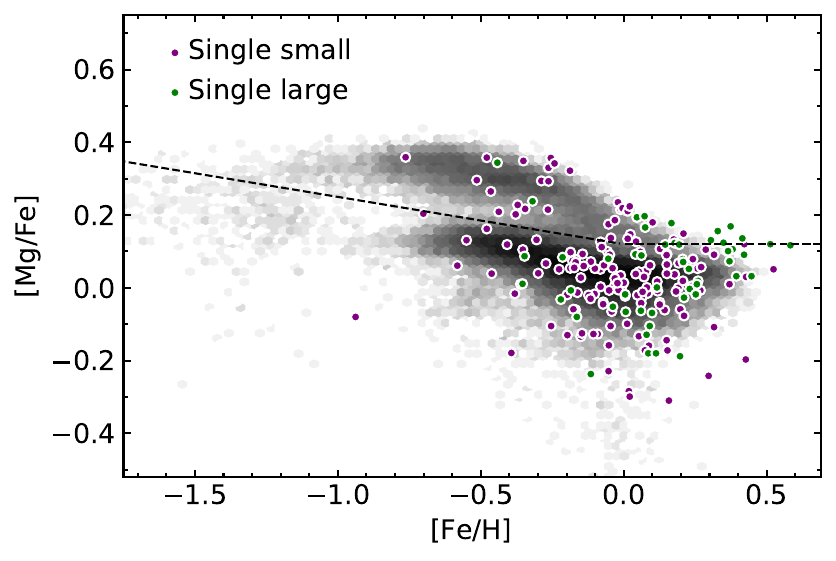}   
    \caption{[Mg/Fe] ratio as a function of [Fe/H] derived in this work (blue circles) compared with the APOGEE data (gray symbols) is shown in the left panel. The dashed line represents the separation between the low-alpha and high-alpha sequences (thin and thick disks according to the division from \cite{weinberg2022ApJS..260...32W}. There is a larger fraction of Single detected Small planets compared to Single detected Large planets in the low-alpha sequence (3.6 times more small planets) when compared to the high-alpha sequence (1.5 times more small planets).} 
    \label{fig:Mg_FeH}
\end{figure*}

To further investigate possible correlations between chemical signatures and planetary properties, in Figure \ref{fig:cdf_feh} we compare the cumulative distribution functions (CDFs) of [Fe/H], [Mg/H], and [Mg/Fe], subdividing our sample into three groups: Singles versus Multis; Small versus Large, and Single Small versus Single Large, based on their planetary architecture. 

For [Fe/H] (second row panel of Figure \ref{fig:cdf_feh}), we find that hosts of Large exoplanets (R${pl} > 4.4$ R$\oplus$) are systematically more metal rich than hosts that only have Small exoplanets; this is also the case for the comparisons between Single Large versus Single Small hosts.
The CDFs clearly show a separation between the populations, for stars with [Fe/H] $>$ 0, and these differences are corroborated by statistical tests, which indicate significance at the $p < 0.01$ level.
This observation aligns with the well-established trend that giant planets preferentially form around metal-rich stars \citep[e.g.,][]{santos2004,fischer2005}. 
On the other hand, the CDFs show only quite subtle differences between Single- and Multi-planet systems (see left panel of second row). The Kolmogorov–Smirnov test (K-S, $p = 0.0197$) and Anderson–Darling test (A-D, $p = 0.0262$) indicate marginally significant differences in the shapes of the distributions at the conventional $\alpha = 0.05$ level. However, these differences do not reach the more stringent $p < 0.01$ threshold, consistent with both \cite{weiss2018} and \cite{Ghezzi2021}, who concluded that the metallicity distributions of the CKS sample for Multis and Singles are indistinguishable when applying this stricter significance criterion. It is worth noting that, in contrast, in an analysis of the metallicities (and oxygen abundances) in a sample of 48 M-dwarf planet-hosting stars, \cite{wanderley2025AJ....170..177W} found that the M dwarfs hosting multi-planet systems display a metallicity distribution function that is shifted significantly to lower values of [M/H] (as well as in [O/H]) when compared to the Single-planet systems (K-S p=0.002).  This difference in the metallicity distributions between Single and Multi systems for the M dwarfs, compared to the more massive FGK stars, may indicate variations in planetary system formation as a function of host-star mass.

Moving to the case of magnesium, the CDFs of [Mg/H] for hosts of Single vs. Multi systems do not show significant differences (left panel of third row), in line with what was shown for [Fe/H]. However, (as also seen for Fe) the CDFs of [Mg/H] for Small vs. Large systems (middle panel of third row) reveal clear differences between the populations, particularly for [Mg/H] $>$ 0, which are confirmed as significant by the statistical tests (K–S and A–D) and the mean [Mg/H] abundances is -0.03 vs. 0.15 dex for Small and Large systems, respectively, similarly to what we find for [Fe/H], mean $<$[Fe/H]$>$ = -0.05 vs. 0.13, for Small versus Large respectively. However, when considering [Mg/Fe] (bottom row panels), the CDFs for Large versus Small, Single Large versus Single Small, and Singles versus Multis largely overlap, and no statistically significant differences are observed. 
In contrast with what was suggested by \cite{adibekyan2012A&A...543A..89A}, our sample of stars hosting Small planets ($R_{pl} < 4.4$ R$_\oplus$) do not tend to have larger values for the [Mg/Fe] ratio when compared with those hosting Large planets.

\subsection{Chemodynamical analysis}

To gain a broader understanding of the sample of exoplanet-hosting stars analyzed in this work, we conducted a dynamical study using astrometric data from Gaia DR3 and chemical information, which are based on high-resolution spectra available for most of the stars. For this purpose, we considered sources with full phase-space information in Gaia DR3 \citep{2023A&A...674A...1G}. To ensure a reliable astrometric solution, we selected sources with \texttt{parallax\_over\_error} $>$ 5, RUWE $<$ 1.4 and \texttt{rv\_expected\_sig\_to\_noise > 5}\footnote{Expected signal to noise ratio in the combination of the spectra used to obtain the radial velocity}. Of the total sample (301 stars), 277 sources met this criterion. Before the analysis, we corrected the parallaxes following the prescription of \cite{2021A&A...649A...4L}. 

\begin{figure}
    \centering
    \includegraphics[width=\linewidth]{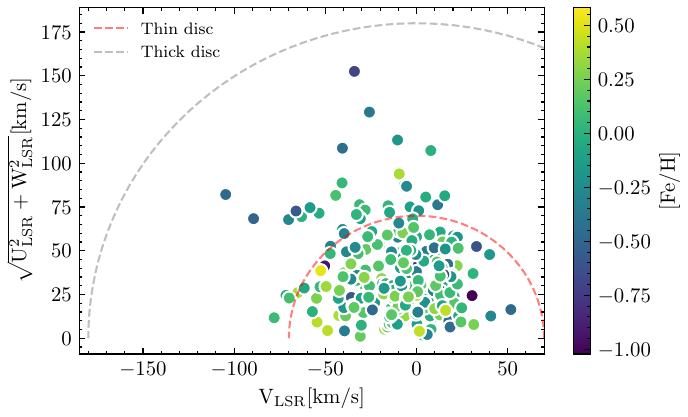}
    \caption{Toomre diagram colored by metallicity. The red dashed line marks the typical velocity boundary of the thin disk (70 km s$^{-1}$), while the region between the red and gray dashed lines (180 km s$^{-1}$) corresponds to the expected kinematic range of the thick disk.} 
    \label{fig:toomre_diagram}
\end{figure}

Kinematic parameters were derived using the Astropy coordinates module \citep{2022ApJ...935..167A}, and orbital parameters were computed using the \texttt{galpy} package \citep{2015ApJS..216...29B}, adopting the MWPotential2014 Galactic potential. Based on the kinematics, all stars in the sample exhibit prograde motion, meaning they move in the direction of Galactic disk rotation. Figure \ref{fig:toomre_diagram} presents the Toomre diagram, color-coded by metallicity, constructed using the Galactic Cartesian velocities (U, V, W) in the local standard of rest (LSR). The red dashed line delineates the typical velocity boundary of the thin disk (70 km s$^{-1}$), while the region between the red and gray dashed lines (180 km s$^{-1}$) corresponds to the expected kinematic space of the thick disk. These boundaries are used as references, since the separation between the two populations is not perfectly well defined.

According to this diagram, none of the stars exhibit halo-like kinematics. Approximately 85\% of the sample displays thin disk kinematics, while the remaining 15\% are consistent with thick disk kinematics. Based on [Mg/Fe], stars classified as thin disk exhibit mean eccentricity and mean $Z_\mathrm{max}$ (maximum height above the Galactic mid plane) values of $0.15 \pm 0.07$ and $0.30 \pm 0.20$ kpc, respectively, whereas sources classified as thick disk show mean eccentricity and mean $Z_\mathrm{max}$ values of $0.33 \pm 0.07$ and $0.50 \pm 0.43$ kpc, in agreement with expectations for these populations.

\subsection{Stellar Activity of the Sample K2 Targets}
\begin{figure*}
    \centering
    \includegraphics[scale=0.45]{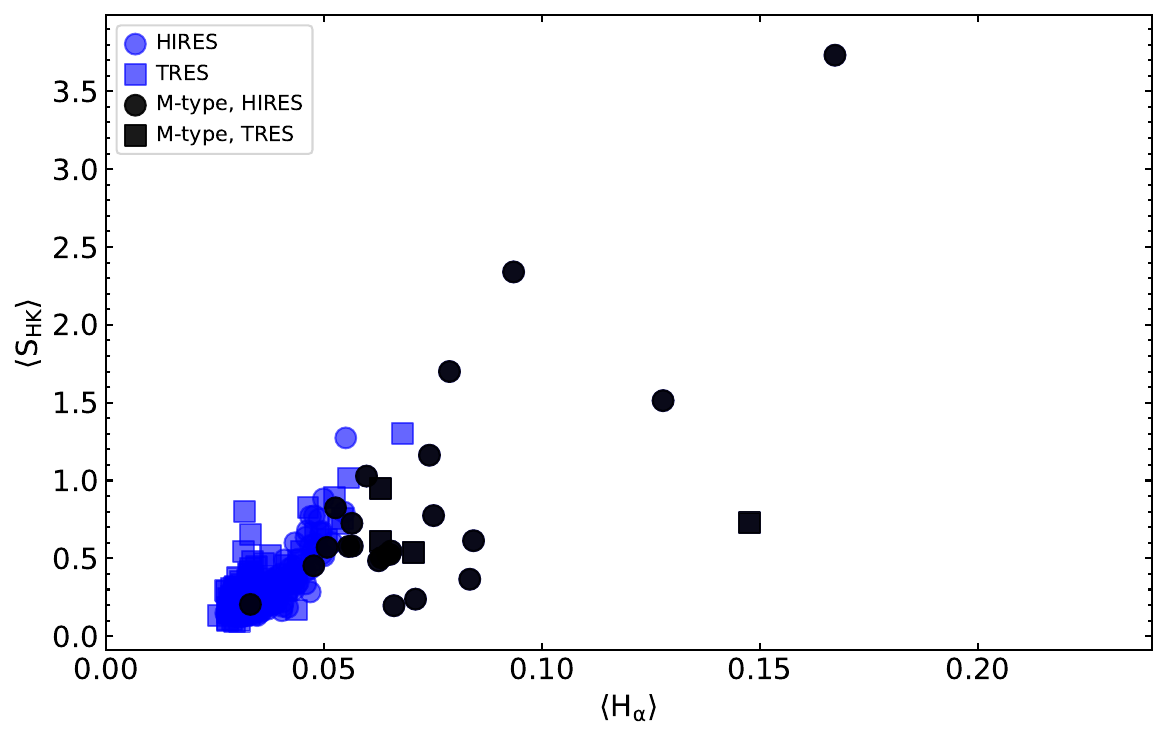}
    \includegraphics[scale=0.45]{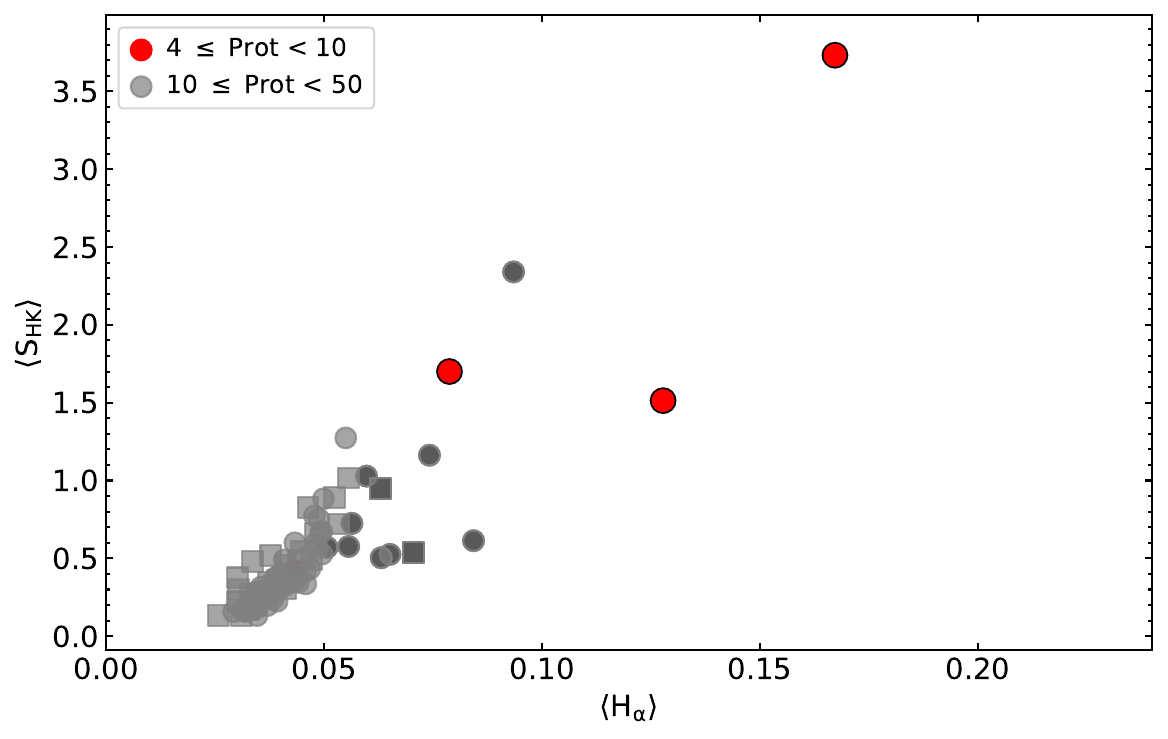}
    \caption{Average S$_{HK}$ as a function of H$_\alpha$ (left panel). Stars with rotational periods whose values are taken from \cite{reinhold2020AA...635A..43R} are shown in the right panel (the black border of the red circle refers to M-type stars).}
    \label{fig:ca_hal}
\end{figure*}

In Figure \ref{fig:ca_hal}, we show the average calcium index as a function of the average H$\alpha$ index for all stars in our sample, separating FGK-type stars (blue circles and squares) from M-type stars (black symbols). 
As noted in Section \ref{sec:data}, only stellar activity indices were measured for M-type stars. In general, as H$\alpha$ increases, S$_{HK}$ also tends to increase, although dispersion is substantial, particularly at higher H$\alpha$ values. Most data points cluster in the region with H$\alpha$ between 0.02 and 0.07, where S$_{HK}$ is below 1, suggesting that the positive correlation may be stronger in this range and becomes more scattered at higher H$\alpha$ values. Interestingly, some stars observed with TRES (blue squares) exhibit nearly constant H$\alpha$ values ($\sim$0.03) while showing a wide range of S$_{HK}$ values, including some of the highest activity levels in the sample. This vertical spread suggests that, in these cases, the Ca II H \& K and H$\alpha$ indices may not trace stellar activity in the same way, potentially due to differences in sensitivity, spectral type dependence, or instrumental calibration. Similar behavior has been reported in previous studies that analyze activity indicators in low-activity or early-type stars \citep[e.g.,][]{gomes2011AA...534A..30G,meunier2022AA...658A..57M}. A possible explanation is the metallicity-dependent bias in Ca II H \& K fluxes identified by \citet{souza2024MNRAS.532..563S}, absent in H$\alpha$, which may explain part of the observed spread when comparing both indicators.

Most M-type stars in our sample display higher levels of stellar activity, with the most active ones reaching S$_{HK}$ and H$\alpha$ index values of about 3.7 and 0.17, respectively. It is well known that a larger fraction of M dwarfs are magnetically active than their more massive Sun-like counterparts (see \cite{wanderley2024ApJ...971..112W,wanderley2024ApJ...975..109W}. 
By dividing stars into active and inactive categories according to their H$\alpha$ EW and spectral type, \citet{kiman2021AJ....161..277K} found that the fraction of active dwarfs decreases with increasing age, and that the shape of this decline depends on spectral type. 
In contrast to FGK dwarf stars, which follow a clearer trend, M dwarfs show greater scatter, consistent with the findings of \citet{gomes2011AA...534A..30G} and \citet{ibanez2023AA...672A..37I} for a sample of M dwarfs.

Stars lose angular momentum and rotational velocity as they evolve, leading to a decrease in stellar activity \citep{skumanich1972ApJ...171..565S,noyes1984ApJ...279..763N}. To study this relationship, we used rotational period measurements from the \cite{reinhold2020AA...635A..43R} catalog. 
We found 87 common stars between our sample and the catalog of \cite{reinhold2020AA...635A..43R}. Most of these stars fall within the range $10 \leq P_{\text{rot}} < 50$ days, while only four stars have periods between $4 \leq P_{\text{rot}} < 10$ days,  as shown in the right panel of Figure \ref{fig:ca_hal}.  
The most active stars (those with higher $S_{HK}$ and H$\alpha$ values) exhibit shorter rotation periods ($4 \leq P_{\text{rot}} < 10$ days), consistent with the presumption that young, rapidly rotating stars tend to be more active. Moreover, an apparent activity threshold emerges, as stars with longer periods ($10 \leq P_{\text{rot}} < 50$ days) do not reach high $S_{HK}$ and H$\alpha$ values. This suggests a decrease in activity over time, likely due to the loss of angular momentum.

\begin{figure*}
    \centering
    \includegraphics[scale=0.49]{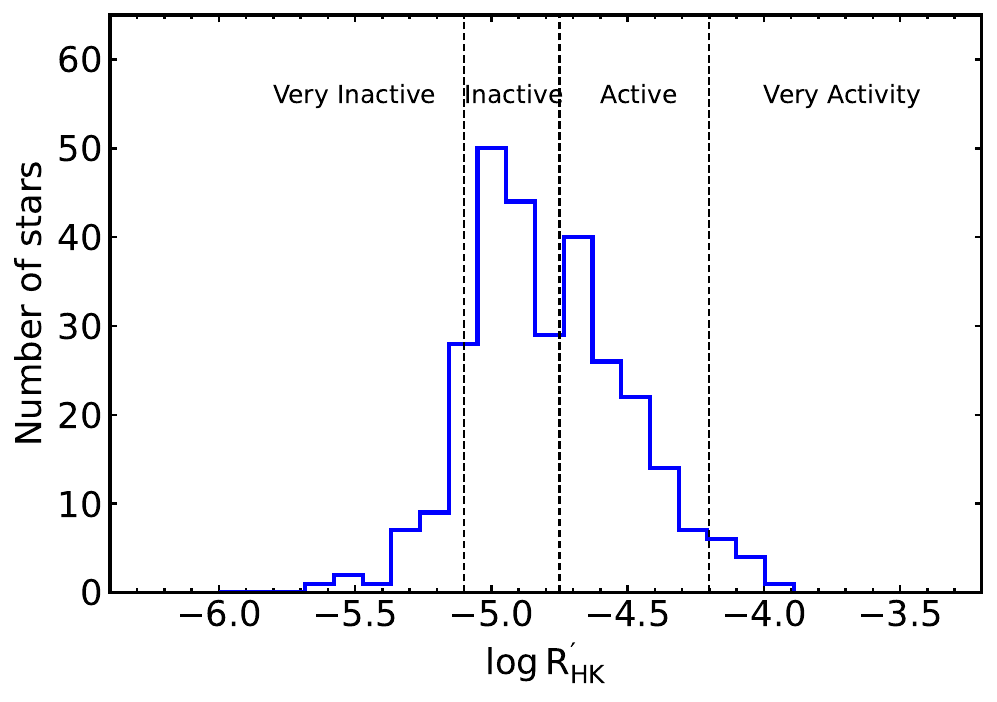}
    \includegraphics[scale=0.49]{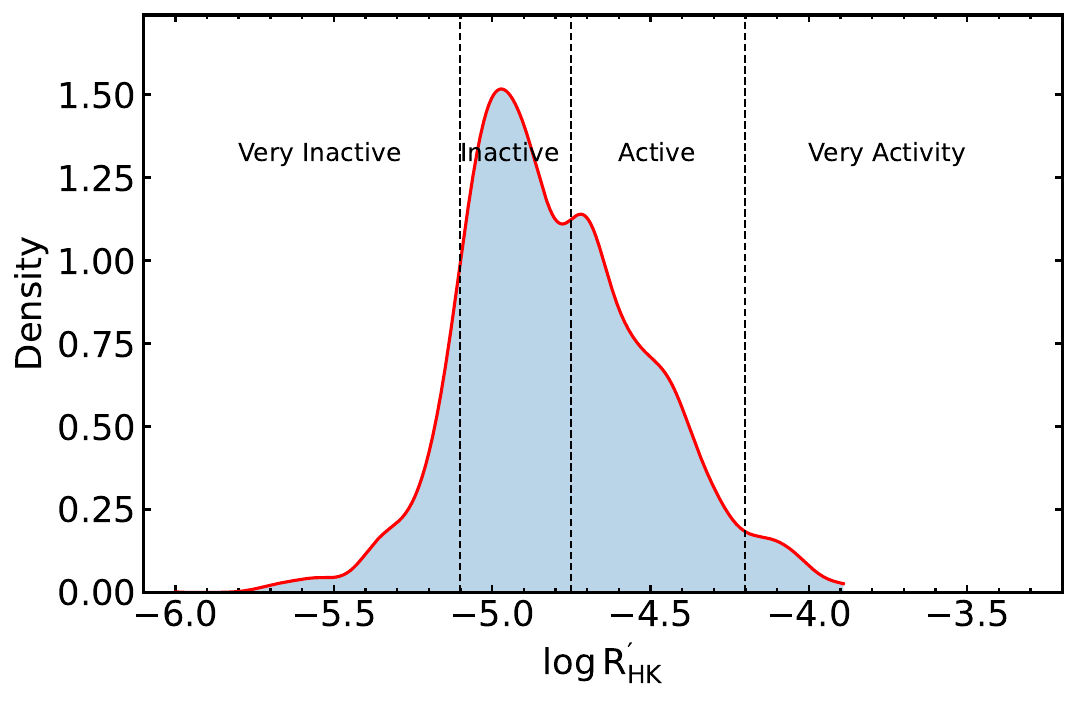}
    \caption{The left panel shows the distribution of $\rm \log R^{\prime}_{HK}$ for the FGK main sequence stars in our sample. In contrast, the right panel presents the $\rm \log R^{\prime}_{HK}$ distribution estimated using Kernel Density Estimation (KDE), which is independent of binning. The gray dashed lines in both panels indicate different activity levels \citep[as identified by][]{henry1996AJ....111..439H}.}
    \label{fig:hist_logRhk}
\end{figure*}

As described in Section~\ref{sec:S_ind}, we corrected the S$_{HK}$ and H$\alpha$ indices for photospheric contributions to obtain the chromospheric activity indicators $\log \mathrm{R}^{\prime}_{HK}$ and $\log \mathrm{I}_{H\alpha}$. This correction enables a more direct comparison of stellar activity levels across different stars, independent of the effects of spectral type.
The distribution of $\log R^{\prime}_{HK}$, shown in Figure~\ref{fig:hist_logRhk}, reveals a bimodal pattern (as previously noted by \citealt{vaughan1978PASP...90..267V}), suggesting two distinct populations of stars with varying levels of chromospheric activity, as shown in our previous study \citep{loaiza2024ApJ...970...53L}. Most of the sample falls into the Inactive category ($-5.1 < \log R^{\prime}_{HK} < -4.75$), indicating that these are predominantly low-activity stars, likely older stars. 
The classification into four activity regimes \citep[Very Inactive, Inactive, Active, and Very Active;][]{henry1996AJ....111..439H} underscores this distinction, with a prominent concentration in the inactive region. A smaller fraction of stars occupies the higher-activity ranges (Active and Very Active, $\log R^{\prime}_{HK} > -4.75$), likely corresponding to younger or magnetically active stars. The two peaks in the distribution are consistent with the notion that stellar activity diminishes over time, causing stars to evolve from an Active to an Inactive phase. This bimodality is more clearly illustrated in the right panel of Figure~\ref{fig:hist_logRhk} through Kernel Density Estimation (KDE), further supporting the existence of two distinct populations with different levels of chromospheric activity. 
No M-type stars are considered in this distribution, as the Vaughan-Preston gap excludes M stars; therefore, the sample includes only FGK stars.
  
\begin{figure*}
    \centering
    \includegraphics[scale=0.55]{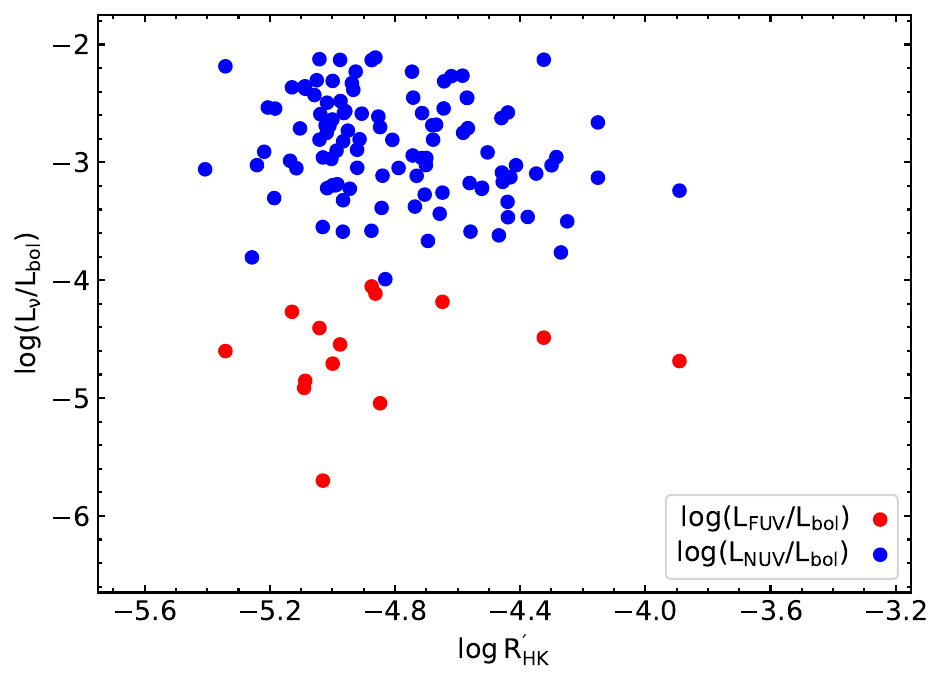}
    \includegraphics[scale=0.55]{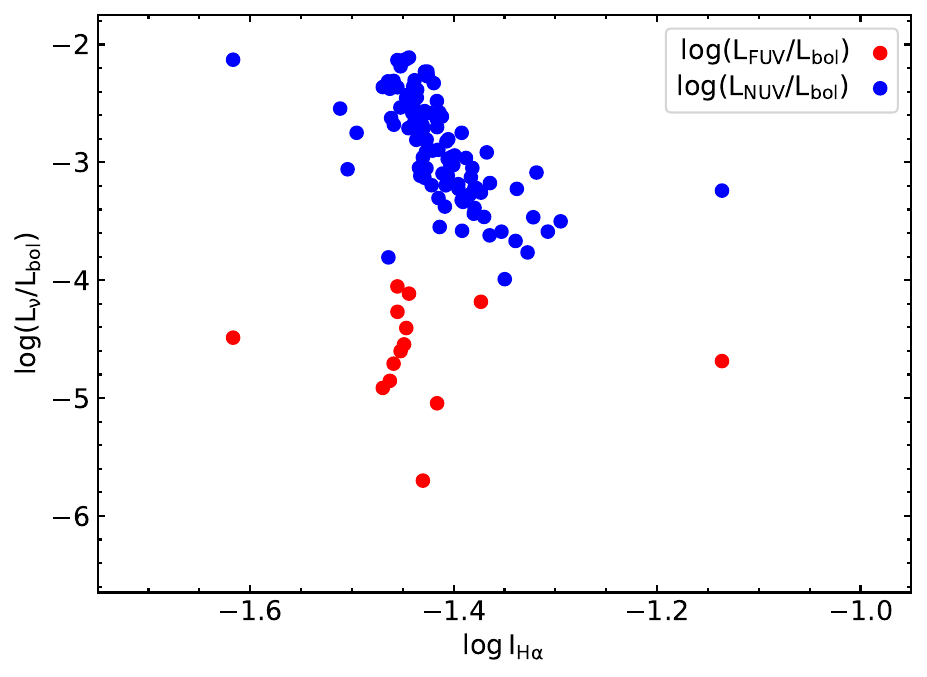}
    \caption{Stellar activity indices ($\log R^\prime _{HK}$ and $\log I_{H\alpha}$) vs. the ratios of near-ultraviolet (NUV) and far-ultraviolet (FUV) to bolometric luminosities for the stars with UV magnitude in GALEX. Red points represent the FUV luminosity fraction, while blue points correspond to the NUV luminosity fraction. FUV emission decreases with lower activity, while NUV remains nearly constant.}
    \label{fig:UV_Rhk_IHa}
\end{figure*}

Figure \ref{fig:UV_Rhk_IHa} shows the far-ultraviolet (FUV) and near-ultraviolet (NUV) to bolometric luminosity ratios plotted versus two activity indicators, the $\log R^{\prime}_{HK}$ and $\log I_{H\alpha}$, for the left and right panels, respectively. 
The FUV and NUV luminosities were determined from FUV and NUV magnitudes obtained from the Galaxy Evolution Explorer \citep[GALEX;][]{martin2005ApJ...619L...1M} ultraviolet (UV) catalog \citep{bianchi2017ApJS..230...24B}. 
In both panels, the FUV luminosity fraction for the fourteen stars with FUV magnitudes in GALEX tends to decrease as the activity indicators decrease, indicating a stronger dependence of FUV emission on stellar activity. The NUV luminosity fraction for 108 stars with NUV magnitudes in GALEX, on the other hand, appears to be constant across the range of activity values, particularly at lower activity levels, reflecting a generally higher contribution from the stellar photosphere. 
The increased variability observed in the FUV emission across the sample may suggest a greater sensitivity to chromospheric or transition region processes.
Overall, this analysis highlights the differences in how FUV and NUV emission relate to stellar activity, with FUV exhibiting a stronger dependence on the activity indicators examined.

We also searched for long-term stellar activity by analyzing stars with more than ten individual visits to identify potential stellar cycles in our sample. We used the S$_{HK}$ index for each stellar observation and performed a frequency analysis using the Lomb-Scargle periodogram \citep{Lomb1976ApSS..39..447L,Scargle1982ApJ...263..835S,VanderPlas2018ApJS..236...16V}. Our results show no well-defined periods, indicating an absence of stellar cycles among the analyzed stars. This may suggest that our current number of observations is insufficient, implying that the timescales of stellar cycles likely exceed our observational coverage (four years for the star with more visits). 

\subsection{Chromospheric Activity and Planetary Radii}
\begin{figure*}
    \centering
    \includegraphics[scale=0.55]{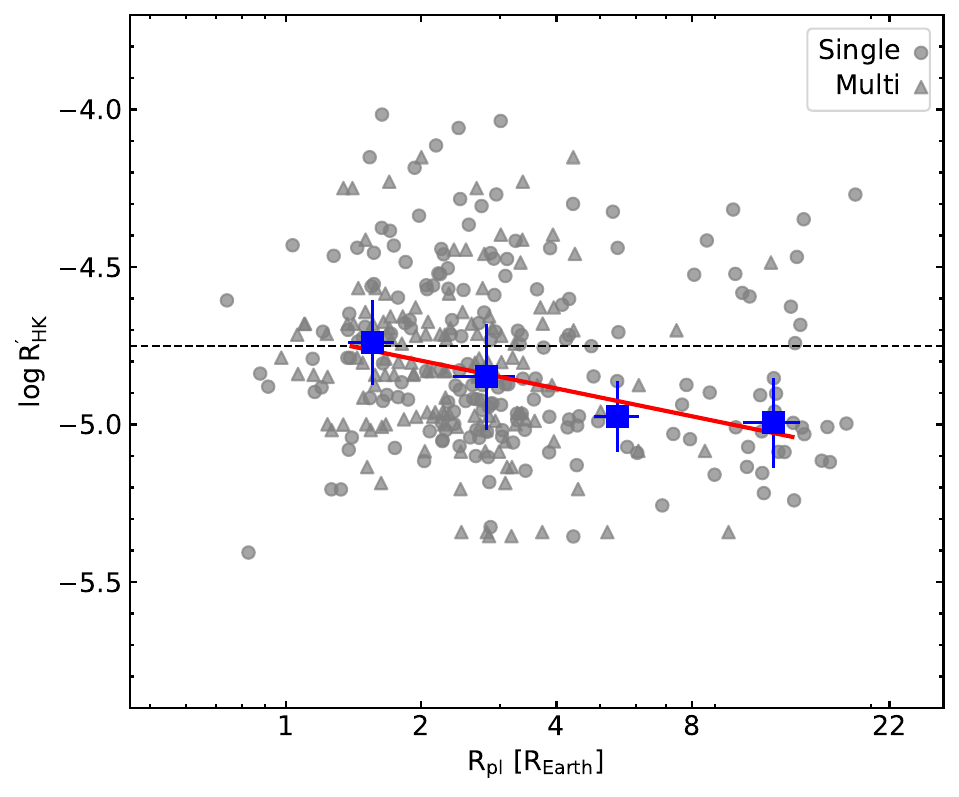}
    \includegraphics[scale=0.55]{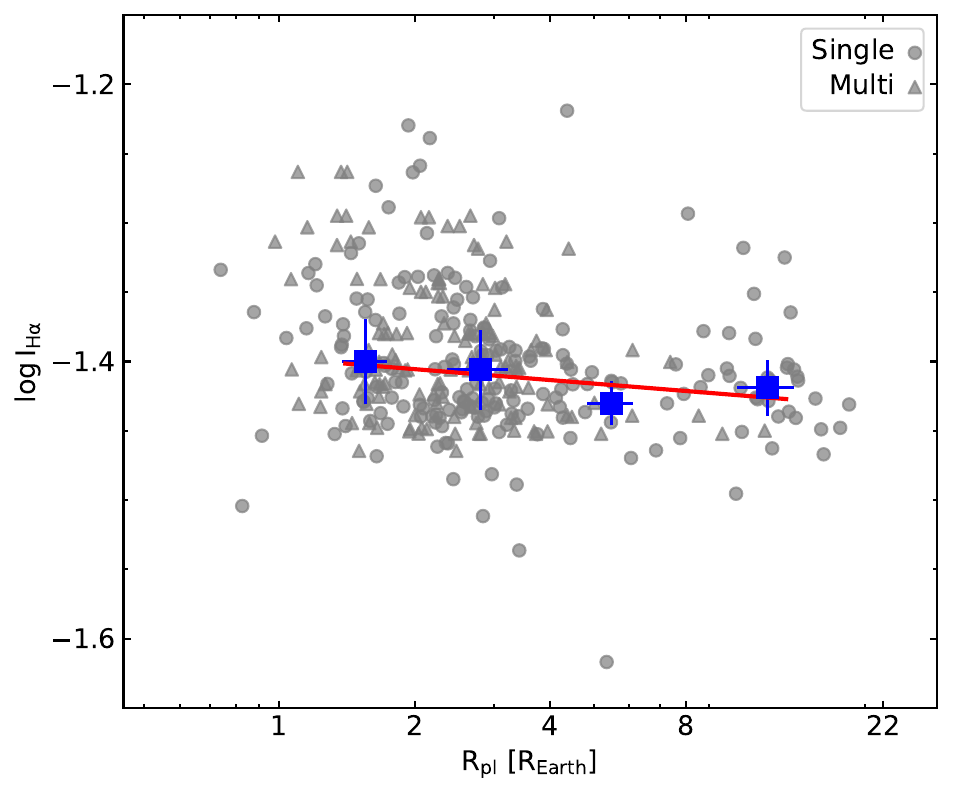}
    \caption{Stellar activity as a function of planetary radius for a sample of 339 exoplanets with host-star stellar activity indices derived in this study. The planetary radii ($R_{\text{pl}}$) are based on stellar radii determined using SB. The blue symbols represent the median host-star stellar activity index for super-Earths ($R_{\text{p}} < 1.9 \, R_{\oplus}$), sub-Neptunes ($1.9 \, R_{\oplus} \leq R_{\text{pl}} < 4.4 \, R_{\oplus}$), sub-Saturns ($4.4 \, R_{\oplus} \leq R_{\text{p}} < 8.0 \, R_{\oplus}$), and Jupiters ($R_{\text{p}} \geq 8.0 \, R_{\oplus}$). The error bars represent the median absolute deviations from these median values.}
    \label{fig:ac_Rpl}
\end{figure*}

Figure \ref{fig:ac_Rpl} illustrates the relationship between stellar activity quantified by $\rm \log R^{\prime}_{HK}$ (left panel) and $\rm \log I_{H\alpha}$ (right panel) as a function of planetary radius (R$_{pl}$) in Earth radii. The circle symbols indicate a single exoplanet, while triangles represent systems with multiple exoplanets, defined as systems with a minimum of two exoplanets. 
Interestingly, within the low-activity range ($\log R^{\prime}_{HK}<-4.75$), in both cases, the data suggest that, as stellar activity decreases, the planetary radius increases. This trend is reflected in the median values (blue squares), which show a slight decrease in activity from smaller to larger planets.

In the case of the calcium index ($\rm \log R^{\prime}_{HK}$, Fig. \ref{fig:ac_Rpl} left panel), the dashed horizontal line in the left plot marks the threshold separating active from inactive stars \citet{henry1996AJ....111..439H}. Interestingly, the median values for different planetary subgroups (super-Earths, sub-Neptunes, sub-Saturns, and Jupiters) show a decreasing trend with increasing planet size, which might suggest that larger planets are preferentially found around less active stars.

A similar pattern is observed for the index $\rm \log I_{H\alpha}$ (Fig. \ref{fig:ac_Rpl}, right panel), where stellar activity also appears to decrease with increasing planetary size, although the data is more scattered. The higher concentration of data points around $\rm \log I_{H\alpha}$ $\sim -1.4$ suggests a possible transition in stellar activity at this level.
This observed trend could indicate that larger planets preferentially form or survive in environments with lower stellar activity. 
In such environments, the high-energy radiation responsible for atmospheric escape is weaker, which would reduce the efficiency of photoevaporation. As a result, planets are more likely to retain their thick gaseous envelopes, leading to larger observed radii \citep{lopez2018MNRAS.479.5303L}. 
On the other hand, the observed trend may be partially influenced by observational biases. For example, small planets are more difficult to detect around highly active stars due to increased stellar noise. In transit photometry, crossings over active regions, such as starspots, can alter the apparent transit depth depending on the contrast with the surrounding photosphere \citep{oshagh2013AA...556A..19O}. Flares and other forms of stellar variability further increase the noise, reducing the detection sensitivity, particularly in active M dwarfs \citep{davenport2016ApJ...829...23D}. 
However, this bias could tend to reduce the number of small planets detected around active stars. Therefore, it cannot account for the observed trend but instead suggests that the underlying correlation between stellar activity and planet size may be even stronger than currently observed.
In radial velocity measurements, starspots, faculae, and plages distort spectral line profiles as the star rotates, inducing RV shifts unrelated to planetary motion \citep{saar1997ApJ...485..319S,queloz2001AA...379..279Q,boisse2011AA...528A...4B}.

Future observations that combine stellar activity indicators with high-precision characterization of planetary atmospheres could help distinguish between true photoevaporation effects and observational biases related to host star activity.
\section{Summary and Conclusions} \label{sec:conclusion}
The stellar parameters, magnesium abundances, radii, masses, and chromospheric activity indices presented in this study were derived from a detailed and uniform analysis of high-resolution optical spectra. This study provides the most extensive and homogeneous spectroscopic characterization of confirmed K2 exoplanet host stars to date. Using a selection of Fe I and Fe II absorption lines, we determined key stellar parameters, including effective temperature, surface gravity, microturbulent velocity, and metallicity. The semi-automated spectroscopic code \texttt{q$^2$} \citep{Ramirez2014} was used to estimate stellar parameters and magnesium abundances for 257 stars. Furthermore, we assessed stellar activity by measuring the Ca II H \& K index and the H$\alpha$ index for 301 stars. Our sample consists of 301 K2 stars with confirmed exoplanets, and the derived stellar radii were subsequently used to determine planetary radii. The main results are summarized below.

\begin{enumerate}
    \item We determined the stellar radii and masses of 257 stars using the Stefan-Boltzmann law, as well as the isochrones method implemented in the codes \texttt{PARAM} and \texttt{isochrones}. For most stars, the derived radii are consistent between these methods. We then determined the planetary radii using transit depths obtained from literature references. 
    The internal precisions of the derived exoplanetary radii were 2.5\%, 2.6\%, and 6.6\% for SB, \texttt{PARAM}, and \texttt{isochrones}, respectively. The radius gap is detected in this sample of K2 planets and is located at approximately $R_{\rm pl} \sim 1.9 R_{\oplus}$ in all three cases.
    \item The cumulative distribution functions of the [Fe/H] and [Mg/H] abundances reveal distinct chemical patterns among planet-hosting stars. Stars hosting large exoplanets ($R_{pl} > 4.4$ R$\oplus$) are systematically more metal-rich than those hosting only small exoplanets, with a statistically significant ($p < 0.01$) difference between populations for [Fe/H] $>$ 0 (mean $<$[Fe/H]$>$ = -0.05 vs. 0.13, respectively), in agreement with the well-established correlation between stellar metallicity and giant planet occurrence. A similar result is obtained for [Mg/H], with stars hosting only small exoplanets or a single detected small exoplanet tending to exhibit lower [Mg/H] values compared to those with giant exoplanets, and stars with a single detected small exoplanet are less enhanced in magnesium than those hosting a single giant exoplanet (mean $<$[Mg/H]$>$ = -0.03 vs 0.15, repectively). For Single- versus Multi-exoplanetary systems we did not find differences for [Fe/H] nor [Mg/H]. These findings suggest that magnesium and iron enrichment play a key role in the formation and evolution of planetary systems, influencing the size of the exoplanets. For the [Mg/Fe] ratio, however, we do not find any significant differences in any of the cases analyzed, suggesting that this ratio does not play an important role in trends with planet size and multiplicity.
    \item The studied K2 sample of exoplanetary hosts is composed of stars from the high- and low-alpha sequences, being both from the thin and thick disks. Our [Mg/Fe] measurements are in good agreement with results from APOGEE DR17, and most of our sample traces the thin-disk sequence. A chemo-dynamical analysis suggests that the majority of stars exhibit consistent behavior in both their kinematics and chemical composition.
    \item We measured stellar activity based on flux in the cores of the Ca II H \& K and H$\alpha$ spectral lines, which serve as proxies for chromospheric activity. Stellar activity using the S$_{HK}$ index, which is defined by the fluxes of the Ca II H \& K lines, was measured for 301 stars using spectra obtained from ExoFop. We calibrated the Mount Wilson index using 36 stars in common with the Mount Wilson Catalog. Photospheric contributions to S$_{HK}$ were subtracted to obtain values of $\log R^{\prime}_{HK}$ using bolometric corrections from \cite{rutten1984AA...130..353R}. We determined H$\alpha$ following the method of \cite{boisse2009AA...495..959B}.
    \item Our analysis confirms that M dwarfs exhibit higher stellar activity than FGK stars, consistent with previous findings \citep{gomes2011AA...534A..30G,wanderley2024ApJ...971..112W}. We observe that the most active stars have shorter rotation periods (4–10 days), while stars with longer periods (10–50 days) show significantly reduced activity. This supports the well-established relationship between stellar rotation and activity, indicating that activity decreases over time due to the loss of angular momentum. Additionally, the greater scatter in activity trends among M dwarfs compared to FGK stars suggests more complex underlying processes affecting their magnetic evolution.
    \item Similar to \cite{henry1996AJ....111..439H} and \cite{gomes2021AA...646A..77G}, who confirmed the Vaughan-Preston gap using a large sample of stars, we also found this feature for K2 exoplanet-hosting sample stars for a larger sample than \cite{loaiza2024ApJ...970...53L}.
    \item Our results indicate that larger planets tend to be found around less active stars,     suggesting a physical link between stellar activity and planetary evolution, such as reduced atmospheric loss via photoevaporation in quieter environments. 
    Within low-activity stars, super-Earths are found around stars with slightly higher activity than those hosting larger planets.
\end{enumerate}

We thank the anonymous referee for their suggestions, which improved the paper.
V.L.T. acknowledges a fellowship 302195/2024-6 of the PCI Program – MCTI and a fellowship 152242/2024-4 of the PDJ - MCTI and CNPq. 
D.S. acknowledges support from the Foundation for Research and Technological Innovation Support of the State of Sergipe (FAPITEC/SE) and the National Council for Scientific and Technological Development (CNPq), under grant numbers 404056/2021-0, 794017/2013, and 444372/2024-5.
S.D. acknowledges CNPq/MCTI for grant 306859/2022-0. This research has used the NASA Exoplanet Archive, which is operated by the California Institute of Technology, under contract with NASA under the Exoplanet Exploration Program.

\bibliography{manuscript}{}
\bibliographystyle{aasjournalv7}
\end{document}